\pdfoutput=1

\documentclass[11pt,twoside,a4paper,cmspaper,final,collab]{cms-tdr}
\def\svnVersion{403846}\def\cmsCernNoTag{CERN-EP/2016-246}\def\cmsCernDate{\today}\def\cmsMessage{Published in the European Physical Journal C as \href{http://dx.doi.org/10.1140/epjc/s10052-017-4787-8}{\doi{10.1140/epjc/s10052-017-4787-8.}}}

\begin{document}\cmsNoteHeader{SUS-15-005}

\hyphenation{had-ron-i-za-tion}
\hyphenation{cal-or-i-me-ter}
\hyphenation{de-vices}
\RCS$HeadURL: svn+ssh://svn.cern.ch/reps/tdr2/papers/SUS-15-005/trunk/SUS-15-005.tex $
\RCS$Id: SUS-15-005.tex 399898 2017-04-20 13:25:28Z bainbrid $

\newcommand{\jets}{\ensuremath{\text{jets}}}
\newcommand{\njet}{\ensuremath{n_{\text{jet}}}\xspace}
\newcommand{\nb}{\ensuremath{n_{\text{b}}}\xspace}
\newcommand{\alphat}{\ensuremath{\alpha_{\mathrm{T}}}\xspace}
\newcommand{\tf}{\ensuremath{\mathcal{T}}\xspace}
\newcommand{\mj}{\ensuremath{\mu + \jets}\xspace}
\newcommand{\mmj}{\ensuremath{\mu\mu + \jets}\xspace}
\newcommand{\mmjpm}{\ensuremath{\mu^\pm\mu^\mp + \jets}\xspace}
\newcommand{\gj}{\ensuremath{\gamma + \jets}\xspace}
\newcommand{\wj}{\ensuremath{\PW + \jets}\xspace}
\newcommand{\wlj}{\ensuremath{\PW (\to  \ell\nu) + \text{jets}}\xspace}
\newcommand{\wmj}{\ensuremath{\PW (\to  \mu\nu) + \text{jets}}\xspace}
\newcommand{\zj}{\ensuremath{\cPZ + \jets}\xspace}
\newcommand{\zmumuj}{\ensuremath{\cPZ (\to  \mu\mu) + \jets}\xspace}
\newcommand{\znunuj}{\ensuremath{\cPZ (\to  \cPgn\cPagn) + \jets}\xspace}
\newcommand{\zmumu}{\ensuremath{\cPZ \to  \mu\mu}\xspace}
\newcommand{\znunu}{\ensuremath{\cPZ \to  \cPgn\cPagn}\xspace}
\newcommand{\ttw}{\ensuremath{\ttbar\PW}\xspace}
\newcommand{\ttz}{\ensuremath{\ttbar\cPZ}\xspace}
\providecommand{\PSl}{\ensuremath{\widetilde{\ell}}\xspace}
\newcommand{\dphi}{\ensuremath{\Delta\phi^{*}_\text{min}}\xspace}
\newcommand{\HTmiss}{\ensuremath{H_{\mathrm{T}}^{\text{miss}}}\xspace}
\newcommand{\mhtmet}{\ensuremath{H_{\mathrm{T}}^{\text{miss}} / E_{\mathrm{T}}^{\text{miss}}}\xspace}
\newcommand{\scalht}{\ensuremath{H_{\mathrm{T}}}\xspace}
\newcommand{\dm}{\ensuremath{\Delta m}\xspace}
\newcommand{\ph}{\ensuremath{\phantom{1}}}
\newcommand{\scalst}{\ensuremath{\mathcal{E}_\mathrm{T}}\xspace}
\newcommand{\dst}{\ensuremath{\Delta\scalst}\xspace}
\newcommand{\bdphi}{\ensuremath{\Delta\phi^{*}_\text{min}}\xspace}
\newcommand{\bdphimod}{\ensuremath{\Delta\phi^{*_{\, 25}}_\text{min}}\xspace}
\newcommand{\cls}{\ensuremath{\mathrm{CL}_\mathrm{s}}\xspace}
\newcommand\T{\rule{0pt}{2.6ex}}
\newcommand\B{\rule[-1.2ex]{0pt}{0pt}}
\newcommand{\NA}{\ensuremath{\text{---}}\xspace}

\newlength\cmsFigWidth
\newlength\cmsFigWidthTwo
\ifthenelse{\boolean{cms@external}}{\setlength\cmsFigWidth{0.49\textwidth}}{\setlength\cmsFigWidth{0.7\textwidth}}
\ifthenelse{\boolean{cms@external}}{\setlength\cmsFigWidthTwo{0.95\textwidth}}{\setlength\cmsFigWidthTwo{0.95\textwidth}}
\ifthenelse{\boolean{cms@external}}{\providecommand{\cmsLeft}{top\xspace}}{\providecommand{\cmsLeft}{left\xspace}}
\ifthenelse{\boolean{cms@external}}{\providecommand{\cmsRight}{bottom\xspace}}{\providecommand{\cmsRight}{right\xspace}}
\ifthenelse{\boolean{cms@external}}{\providecommand{\cmsTable}[1]{#1}}{\providecommand{\cmsTable}[1]{\resizebox{\textwidth}{!}{#1}}}

\newlength\cmsTableLabelSkip
\setlength\cmsTableLabelSkip{-1.2ex}

\hyphenation{brems-strah-lung}
\cmsNoteHeader{SUS-15-005}

\title{A search for new phenomena in pp collisions at $\sqrt{s} =
  13\TeV$ in final states with missing transverse momentum and at
  least one jet using the \alphat variable}
\titlerunning{Search for new phenomena at 13\TeV with \ETmiss and jets using \alphat}

\date{\today}

\abstract{ A search for new phenomena is performed in final states
  containing one or more jets and an imbalance in transverse momentum
  in pp collisions at a centre-of-mass energy of 13\TeV. The analysed
  data sample, recorded with the CMS detector at the CERN LHC,
  corresponds to an integrated luminosity of 2.3\fbinv.  Several
  kinematic variables are employed to suppress the dominant
  background, multijet production, as well as to discriminate between
  other standard model and new physics processes. The search provides
  sensitivity to a broad range of new-physics models that yield a
  stable weakly interacting massive particle. The number of observed
  candidate events is found to agree with the expected contributions
  from standard model processes, and the result is interpreted in the
  mass parameter space of fourteen simplified supersymmetric models
  that assume the pair production of gluinos or squarks and a range of
  decay modes. For models that assume gluino pair production, masses
  up to 1575 and 975\GeV are excluded for gluinos and neutralinos,
  respectively. For models involving the pair production of top
  squarks and compressed mass spectra, top squark masses up to 400\GeV
  are excluded.}

\hypersetup{
  pdfauthor={CMS Collaboration},
  pdftitle={A search for new phenomena in pp collisions at sqrt(s) = 13 TeV in
  final states with missing transverse momentum and at least one jet
  using the alphaT variable},
  pdfsubject={CMS},
  pdfkeywords={CMS, jets, missing transverse momentum, supersymmetry,
    dark matter, AlphaT},
}

\maketitle

\section{Introduction}
\label{sec:introduction}

The standard model (SM) of particle physics is successful in
describing a wide range of phenomena, although it is widely believed
to be only an effective approximation of a more complete theory that
supersedes it at a higher energy scale. Supersymmetry
(SUSY)~\cite{ref:SUSY-1, ref:SUSY0, ref:SUSY3, ref:SUSY1} is a
modification to the SM that extends its underlying space-time symmetry
group. For each boson (fermion) in the SM, a fermionic (bosonic)
superpartner, which differs in spin by one-half unit, is introduced.

Experimentally, SUSY is testable through the prediction of an
extensive array of new observable states (of unknown
masses)~\cite{ref:SUSY4, ref:SUSY2}. In the minimal supersymmetric
extension to the SM~\cite{ref:SUSY2}, the gluinos $\PSg$, light- and
heavy-flavour squarks $\PSQ, \PSQb, \PSQt$, and sleptons $\PSl$ are,
respectively, the superpartners to gluons, quarks, and leptons. An
extended Higgs sector is also predicted, as well as four neutralino
$\PSGcz_{1,2,3,4}$ and two chargino $\PSGcpm_{1,2}$ states that arise
from mixing between the higgsino and gaugino states, which are the
superpartners of the Higgs and electroweak gauge bosons.  The
assumption of $R$-parity conservation~\cite{Farrar:1978xj} has
important consequences for cosmology and collider
phenomenology. Supersymmetric particles are expected to be produced in
pairs at the LHC, with heavy coloured states decaying, potentially via
intermediate SUSY states, to the stable lightest SUSY particle
(LSP). The LSP is generally assumed to be the $\PSGczDo$, which is
weakly interacting and massive. This SUSY particle is considered to be
a candidate for dark matter (DM)~\cite{Jungman:1995df}, the existence
of which is supported by astrophysical
data~\cite{1674-1137-38-9-090001}.  Hence, a characteristic signature
of R-parity-conserving coloured SUSY production at the LHC is a final
state containing an abundance of jets, possibly originating from top
or bottom quarks, accompanied by a significant transverse momentum
imbalance, \ptvecmiss.

The proposed supersymmetric extension of the SM is also compelling
from a theoretical perspective, as the addition of superpartners to SM
particles can modify the running of the gauge coupling constants such
that their unification can be achieved at a high energy
scale~\cite{Dimopoulos:1981yj, Ibanez:1981yh, Marciano:1981un}. A more
topical perspective, given the recently discovered Higgs
boson~\cite{ref:atlashiggsdiscovery, ref:cmshiggsdiscovery,
  ref:cmshiggsdiscoverylong}, is the possibility that scale-dependent
radiative corrections to the Higgs boson mass from loop processes can
be largely cancelled through the introduction of superpartners, thus
alleviating the gauge hierarchy problem~\cite{ref:hierarchy1,
  ref:hierarchy2}. Alternatively, these radiative corrections can be
accommodated through an extreme level of fine tuning of the bare Higgs
boson mass. A ``natural'' solution from SUSY, with minimal
fine-tuning, implies that the masses of the $\PSGczDo$,
third-generation squarks, and the gluino are at or near the
electroweak scale~\cite{ref:barbierinsusy}.

The lack of evidence to date for SUSY has also focused attention on
regions of the natural parameter space with sparse experimental
coverage, such as phenomenologically well motivated models for which
both the $\PSQt$ and the $\PSGczDo$ are light and nearly degenerate in
mass~\cite{Boehm:1999bj, Boehm:1999tr, Balazs:2004bu, Martin:2007gf,
  Martin:2007hn, Carena:2008mj, Delgado:2012eu, Grober:2014aha,
  Grober:2015fia}. This class of models, with ``compressed'' mass
spectra, typically yield SM particles with low transverse momenta
(\pt) from the decays of SUSY particles. Hence, searches rely on the
associated production of jets, often resulting from initial-state
radiation (ISR), to achieve experimental acceptance.

This paper presents an inclusive search for new-physics phenomena in
hadronic final states with one or more energetic jets and an imbalance
in \ptvecmiss, performed in proton-proton (pp) collisions at a
centre-of-mass energy $\sqrt{s} = 13\TeV$. The analysed data sample
corresponds to an integrated luminosity of $2.3 \pm 0.1\fbinv$
collected by the CMS experiment. Earlier searches using the same
technique have been performed in pp collisions at both $\sqrt{s} =
7$~\cite{RA1Paper, RA1Paper2011, RA1Paper2011FULL} and
$8\TeV$~\cite{RA1Paper2012, RA1Parked} by the CMS Collaboration.
The increase in the centre-of-mass energy of the LHC, from $\sqrt{s} =
8$ to 13\TeV, provides a unique opportunity to search for the
characteristic signatures of new physics at the \TeV scale. For
example, the increase in $\sqrt{s}$ leads to a factor $\gtrsim$35
increase in the parton luminosity~\cite{susynlo} for the pair
production of coloured SUSY particles, each of mass 1.5\TeV, which
were beyond the reach of searches performed at $\sqrt{s} = 8\TeV$ by
the ATLAS~\cite{Aad:2015iea, Aad:2015pfx} (and references therein) and
CMS~\cite{CMS:2014dpa, Khachatryan:2015vra, Khachatryan:2016oia,
  Chatrchyan:2013wxa, Chatrchyan:2014lfa, Khachatryan:2015pwa,
  Khachatryan:2015wza, Khachatryan:2016zcu} Collaborations. Several
searches in this final state, interpreted within the context of SUSY,
have already provided results with the first data at this new energy
frontier~\cite{Aad:2016jxj, Aaboud:2016tnv, Aaboud:2016zdn,
  Aad:2016eki, Aaboud:2016nwl, Khachatryan:2016kdk, cms-13}.

Two important features of this search are the application of selection
criteria with low thresholds, in order to maximise signal acceptance,
and the categorisation of candidate signal events according to
multiple discriminating variables for optimal signal extraction over a
broad range of models. The search is based on an examination of the
number of reconstructed jets per event, the number of these jets
identified as originating from bottom quarks, and the scalar and
vector \pt sums of these jets. These variables provide sensitivity to
the different production mechanisms (squark-squark, squark-gluino, and
gluino-gluino) of massive coloured SUSY particles at hadron colliders,
third-generation squark signatures, and both large and small mass
splittings between the parent SUSY particle and the LSP. However, the
search is sufficiently generic and inclusive to provide sensitivity to
a wide range of SUSY and non-SUSY models that postulate the existence
of a stable, only weakly interacting, massive particle. In addition to
the jets+\ptvecmiss topology, the search considers final states
containing a ``monojet'' topology, which is expected to improve the
sensitivity to DM particle production in pp
collisions~\cite{Fox:2012ee, Buchmueller:2015eea}.

The dominant background process for a search in all-jet final states
is multijet production, a manifestation of quantum chromodynamics
(QCD). An accurate estimate of this background is difficult to
achieve, given the lack of precise theoretical predictions for the
multijet production cross section and kinematic properties.
Hence, this search adopts a strategy that employs several variables to
reduce the multijet contribution to a low level with respect to other
SM backgrounds, rather than estimate a significant contribution with
high precision.

The search is built around two variables that are designed to provide
robust discrimination against multijet events. A dimensionless
kinematic variable \alphat~\cite{Randall:2008rw, RA1Paper} is
constructed from jet-based quantities and provides discrimination
between genuine sources of \ptvecmiss from stable, weakly interacting
particles such as neutrinos or neutralinos that escape the detector,
and instrumental sources such as the mismeasurements of jet
energies. The \bdphi~\cite{RA1Paper} variable exploits azimuthal
angular information and also provides strong rejection power against
multijet events, including rare energetic events in which neutrinos
carry a significant fraction of the energy of a jet due to
semileptonic decays of heavy-flavour mesons. Very restrictive
requirements on the \alphat and \dphi variables are employed in this
search to ensure a low level of contamination from the multijet
background.

The organisation of this paper is as
follows. Sections~\ref{sec:detector} and~\ref{sec:simulation}
describe, respectively, the CMS apparatus and the simulated event
samples. Sections~\ref{sec:event_reconstruction}
and~\ref{sec:event_selection} describe the event reconstruction and
selection criteria used to identify candidate signal events and
control region samples. Section~\ref{sec:backgrounds} provides details
on the estimation of the multijet and all other SM
backgrounds. Finally, the search results and interpretations, in terms
of simplified SUSY models, are described in
Sections~\ref{sec:result} and~\ref{sec:interpretations}, and
summarised in Section~\ref{sec:summary}.

\section{The CMS detector}
\label{sec:detector}

The central feature of the CMS apparatus is a superconducting solenoid
of 6\unit{m} internal diameter, providing an axial magnetic field of
3.8\unit{T}. The bore of the solenoid is instrumented with several
particle detection systems. A silicon pixel and strip tracker measures
charged particles within the pseudorapidity range $\abs{\eta} < 2.5$.
A lead tungstate crystal electromagnetic calorimeter (ECAL), and a
brass and scintillator hadron calorimeter (HCAL), each composed of a
barrel and two endcap sections, extend over a range $\abs{\eta} <
3.0$.
Outside the bore of the solenoid, forward calorimeters extend the
coverage to $\abs{\eta} < 5.0$, and muons are measured within
$\abs{\eta} < 2.4$ by gas-ionisation detectors embedded in the steel
flux-return yoke outside the solenoid. A two-tier trigger system
selects pp collision events of interest. The first level of the
trigger system, composed of custom hardware processors, uses
information from the calorimeters and muon detectors to select the
most interesting events in a fixed time interval of less than
4\mus. The high-level trigger processor farm further decreases the
event rate from around 100\unit{kHz} to less than 1\unit{kHz}, before
data storage. The CMS detector is nearly hermetic, which allows for
momentum balance measurements in the plane transverse to the beam
axis. A more detailed description of the CMS detector, together with a
definition of the coordinate system used and the relevant kinematic
variables, can be found in Ref.~\cite{Chatrchyan:2008zzk}.

\section{Simulated event samples}
\label{sec:simulation}

The search relies on multiple event samples, in data or generated from
Monte Carlo (MC) simulations, to estimate the contributions from SM
backgrounds, as described in Section~\ref{sec:event_selection}.
The SM backgrounds for the search are QCD multijet, top
quark-antiquark (\ttbar), and single top production, and the
associated production of jets and a vector boson (W,
\znunu). Residual contributions from other processes, such as WW, WZ,
ZZ (diboson) production and the associated production of \ttbar and a
W or Z boson, are also considered. Other processes, such as Drell--Yan
($\cPq\bar{\cPq} \to  \PZ/\gamma^*  \to
\ell^+\ell^-$) and \gj production, are also relevant for some control
regions, defined in Section~\ref{sec:control_regions}.

The {\MADGRAPH{}5\_a\MCATNLO} 2.2.2~\cite{Alwall2014} event generator
code is used at leading-order (LO) accuracy to produce samples of \wj,
\zj, \gj, \ttbar, and multijet events. The same code is used at
next-to-leading-order (NLO) accuracy to generate samples of single top
quarks (both $s$- and $t$-channel production), WZ, ZZ, \ttw, and \ttz
events. The NLO \POWHEG v2~\cite{powheg, powheg_top_Wt} generator is
used to describe WW events and the tW-channel production of single top
quark events. The simulated samples are normalised according to
production cross sections that are calculated with NLO and next-to-NLO
precision~\cite{Alwall2014, wphys, fewz, wwxs, top++, nlotop,
  powheg_top_Wt}, or with LO precision in the case of multijet and \gj
production.  The \GEANTfour~\cite{geant} package is used to simulate
the detector response.

{\tolerance=1200
Event samples for signal models involving gluino or squark pair
production in association with up to two additional partons are
generated at LO with {\MADGRAPH{}5\_a\MCATNLO}, and the
decay of the SUSY particles is performed with \PYTHIA
8.205~\cite{pythia}. Inclusive, process-dependent, signal production
cross sections are calculated with NLO plus
next-to-leading-logarithmic (NLL) accuracy~\cite{Beenakker:1996ch,
  PhysRevLett.102.111802, PhysRevD.80.095004, 1126-6708-2009-12-041,
  doi:10.1142/S0217751X11053560, susynlo}. The theoretical systematic
uncertainties are typically dominated by the parton density function
(PDF) uncertainties, evaluated using the
CTEQ6.6~\cite{Nadolsky:2008zw} and MSTW2008~\cite{Martin:2009iq} PDFs.
The detector response for signal models is provided by the CMS fast
simulation package~\cite{fastsim}.
\par}

The {NNPDF}3.0 LO and NLO~\cite{nnpdf} parton distribution
functions (PDFs) are used, respectively, with the LO and NLO
generators described above. The \PYTHIA program with the
CUETP8M1 underlying event tune~\cite{Khachatryan:2015pea} is used
to describe parton showering and hadronisation for all simulated
samples. To model the effects of multiple pp collisions within the
same or neighbouring bunch crossings (pileup), all simulated events
are generated with a nominal distribution of pp interactions per bunch
crossing and then reweighted to match the pileup distribution as
measured in data. On average, approximately fifteen different pp
collisions, identifiable via their primary interaction vertex, are
reconstructed per event. Finally, (near-unity) corrections to the
normalisation of the simulated samples for the \gj, \wmj, \ttbar,
\zmumuj, and \znunuj processes are derived using data sidebands to the
control regions.

\section{Event reconstruction}
\label{sec:event_reconstruction}

Global event reconstruction is provided by the particle flow (PF)
algorithm~\cite{CMS-PAS-PFT-09-001,CMS-PAS-PFT-10-001}, designed to
identify each particle using an optimised combination of information
from all detector systems. In this process, the identification of the
particle type (photon, electron, muon, charged hadron, neutral hadron)
plays an important role in the determination of the particle direction
and energy.

Among the vertices reconstructed within 24\,(2)\unit{cm} of the
detector centre parallel (perpendicular) to the beam axis, the primary
vertex (PV) is assigned to be the one with the largest sum of charged
particle (track) $\pt^2$ values.
Charged-particle tracks associated with reconstructed vertices from
pileup events are not considered by the PF algorithm as part of the
global event reconstruction.

Photon candidates~\cite{CMS:EGM-14-001} are identified as ECAL energy
clusters not linked to the extrapolation of any track to the ECAL. The
energy of photons is directly obtained from the ECAL measurement,
corrected for contributions from pileup events.  Various
quality-related criteria must be satisfied in order to identify
photons with high efficiency while minimising the misidentification of
electrons and associated bremsstrahlung, jets, or ECAL noise as
photons. The criteria include the following: the shower shape of the
energy deposition in the ECAL must be consistent with that expected
from a photon, the energy detected in the HCAL behind the photon
shower must not exceed 5\% of the photon energy, and no matched hits
in the pixel tracker must be found.

Electron candidates~\cite{Khachatryan:2015hwa} are identified as a
track associated with an ECAL cluster compatible with
the track trajectory, as well as additional ECAL energy clusters from
potential bremsstrahlung photons emitted as the electron traverses
material of the silicon tracker. The energy of electrons is determined
from a combination of the track momentum at the main interaction
vertex, the corresponding ECAL cluster energy, and the energy sum of
all bremsstrahlung photons associated with the track. The quality
criteria required for electrons are similar to those for photons, with
regards to the ECAL shower shape and the relative contributions to the
total energy deposited in the ECAL and HCAL. Additional requirements
are also made on the associated track, which consider the track
quality, energy-momentum matching, and compatibility with the PV in
terms of the transverse and longitudinal impact parameters.

Muon candidates~\cite{Chatrchyan:2012xi} are identified as a track in the
silicon tracker consistent with either a track or several hits in the
muon system. The track and hit parameters must satisfy various
quality-related criteria, described in Ref.~\cite{Chatrchyan:2012xi}.
The energy of muons is obtained from the corresponding track momentum.

Charged hadrons are identified as tracks not
classified as either electrons or muons.
The energy of charged hadrons is determined from a combination of the
track momentum and the corresponding ECAL and HCAL energies, corrected
for contributions from pileup events and the response function of the
calorimeters to hadronic showers. Neutral hadrons are identified as
HCAL energy clusters not linked to any charged-hadron trajectory, or
as ECAL and HCAL energy excesses with respect to the expected
charged-hadron energy deposit.  The energy of neutral hadrons is
obtained from the corresponding corrected ECAL and HCAL energy.

Photons are required to be isolated from other activity in the event,
such as charged and neutral hadrons, within a cone $\Delta R =
\sqrt{\smash[b]{(\Delta\phi)^2 + (\Delta\eta)^2}} = 0.3$ around the photon
trajectory, corrected for contributions from pileup events and the
photon itself. Electrons and muons are also required to be isolated
from other reconstructed particles in the event, primarily to suppress
background contributions from semileptonic heavy-flavour decays in
multijet events. The isolation $I^\text{mini}_\text{rel}$ is defined
as the scalar \pt sum of all charged and neutral hadrons, and photons,
within a cone around the lepton direction, divided by the lepton
\pt. The ``mini'' cone radius is dependent on the lepton \pt,
primarily to identify with high efficiency the collimated daughter
particles of semileptonically decaying Lorentz-boosted top quarks,
according to the following: $R = 0.2$ and 0.05 for, respectively, $\pt
< 50\GeV$ and $\pt > 200\GeV$, and $R = 10\GeV / \pt$ for $50 < \pt <
200\GeV$. The variable $I^\text{mini}_\text{rel}$ excludes
contributions from the lepton itself and pileup events. The isolation
for electrons and muons is required to satisfy, respectively,
$I^\text{mini}_\text{rel} < 0.1$ and 0.2 for the signal region and
nonleptonic control sample selection criteria.  A tighter definition
of muon isolation $I^{\mu}_\text{rel}$ is used for the definition of
control regions that are required to contain at least one muon. The
variable $I^{\mu}_\text{rel}$ is determined identically to
$I^\text{mini}_\text{rel}$ except that a cone of fixed radius $R =
0.4$ is assumed.

Electron and muon candidates identified by the PF algorithm that do
not satisfy the quality criteria or the $I^\text{mini}_\text{rel}$
isolation requirements described above, as well as charged hadrons,
are collectively labelled as ``single isolated tracks'' if they are
isolated from neighbouring tracks associated to the
PV. The isolation $I^\text{track}_\text{rel}$ is defined as the scalar
\pt sum of tracks (excluding the track under
consideration) within a cone $\Delta R < 0.3$ around the track
direction, divided by the track \pt. The requirement
$I^\text{track}_\text{rel} < 0.1$ is imposed.

Jets are clustered from the PF candidate particles with the infrared-
and collinear-safe anti-$k_t$ algorithm~\cite{antikt}, operated with a
distance parameter of 0.4. The jet momentum is determined as the
vectorial sum of all particle momenta in the jet, and is found in the
simulation to be within 5 to 10\% of its true momentum over the whole
\pt spectrum and detector acceptance. Jet energy corrections, to
account for pileup~\cite{pileup} and to establish a uniform relative
response in $\eta$ and a calibrated absolute response in \pt, are
derived from the simulation, and are confirmed with in situ
measurements using the energy balance in dijet and photon+jet
events~\cite{Chatrchyan:2011ds}. The jet energy resolution is
typically 15\% at 10\GeV, 8\% at 100\GeV, and 4\% at 1\TeV, compared
to about 40\%, 12\%, and 5\% obtained when the calorimeters alone are
used for jet clustering.
All jets are required to satisfy loose requirements on the relative
composition of their particle constituents to reject noise in the
calorimeter systems or failures in event reconstruction.

Jets are identified as originating from b quarks using the combined
secondary vertex algorithm~\cite{CMS-PAS-BTV-12-001}. Control regions
in data~\cite{bjets} are used to measure the probability of correctly
identifying jets as originating from b quarks (b tagging efficiency),
and the probability of misidentifying jets originating from
light-flavour partons (u, d, s quarks or gluons) or a charm quark as a
b-tagged jet (the light-flavour and charm mistag probabilities). A
working point is employed that yields a b tagging efficiency of 65\%,
and charm and light-flavour mistag probabilities of approximately 12
and 1\%, respectively, for jets with \pt that is typical of \ttbar
events.

An estimator of \ptvecmiss is given by the projection on the plane
perpendicular to the beams of the negative vector sum of the momenta
of all candidate particles in an event~\cite{cms-met}, as determined
by the PF algorithm. Its magnitude is referred to as \ETmiss.

\section{Event selection}
\label{sec:event_selection}

The kinematic selection criteria used to define the signal region,
containing a sample of candidate signal events, as well as a number of
control regions in data, are described below. The criteria are based
on the particle candidates defined by the event reconstruction
algorithms described in Section~\ref{sec:event_reconstruction}.

\subsection{Common preselection criteria}
\label{sec:preselection}

A number of beam- and detector-related effects, such as beam halo,
reconstruction failures, spurious detector noise, or event
misreconstruction due to detector inefficiencies, can lead to events
with anomalous levels of activity. These rare events, which can
exhibit large values of \ETmiss, are rejected with high efficiency by
applying a range of dedicated vetoes~\cite{1748-0221-5-03-T03014,
  cms-met}.

In order to suppress SM processes with genuine \ptvecmiss from
neutrinos, events containing an isolated electron or muon that
satisfies $\pt > 10\GeV$ and $\abs{\eta} < 2.5$ are vetoed. Events
containing an isolated photon with $\pt > 25\GeV$ and $\abs{\eta} <
2.5$ are also vetoed, in order to select only multijet final
states. Furthermore, events containing a single isolated track
satisfying $\pt > 10\GeV$ and $\abs{\eta} < 2.5$ are vetoed in order
to reduce the background contribution from final states containing
hadronically decaying tau leptons.

Each jet $\mathrm{j}_i$ considered by this search is required to satisfy
$\pt^{\mathrm{j}_i} > 40\GeV$ and $|\eta^{\mathrm{j}_i}| < 3$. The number
of jets within this experimental acceptance is labelled henceforth as
\njet. The highest \pt jet in the event is required to have
$\pt^{\mathrm{j_1}} > 100\GeV$ and $\abs{\eta^{\mathrm{j_1}}} < 2.5$. The
second-highest \pt jet in the event is used to categorise events, as
described in Section~\ref{sec:categorisation}. If the jet satisfies
$\pt^{\mathrm{j_2}} > 100\GeV$, then this category of events is labelled
``symmetric'' and targets primarily topologies resulting from
pair-produced SUSY particles.
If the jet satisfies $40 < \pt^{\mathrm{j_2}} < 100\GeV$ then the event
is labelled as ``asymmetric,'' and if there exists no second jet with
$\pt^{\mathrm{j_2}} > 40\GeV$, the event is labelled monojet. The
asymmetric and monojet topologies target models involving the direct
production of stable, weakly interacting, massive particles.
The mass scale of the physics processes being probed is characterised
by the scalar \pt sum of the jets, defined as $\scalht =
\sum_{i=1}^{\njet} \pt^{\,\mathrm{j}_i}$.
The magnitude of the vector \ptvec sum of these jets, defined by
$\HTmiss = \abs{\sum_{{i}=1}^{\njet} \ptvec^{\,\mathrm{j}_i}}$, is
used to identify events with significant imbalance in
\ptvecmiss. Events are vetoed if any jet satisfies $\pt > 40\GeV$ and
$\abs{\eta} > 3$ to ensure that jets reconstructed in the forward regions
of the detector do not contribute significantly to \HTmiss.

The dimensionless variable $\HTmiss / \ETmiss$ is used to remove
events that contain several jets with transverse momenta below the jet
\pt thresholds but an appreciable vector \pt sum so as to contribute
significantly to \HTmiss relative to \ETmiss. This background is
typical of multijet events, which is suppressed by requiring $\HTmiss
/ \ETmiss < 1.25$. The requirement is imposed as part of the common
preselection criteria used to define all control samples to minimise
potential systematic biases associated with the simulation modelling
for this variable. A high efficiency is maintained for SM or
new-physics processes that produce unobserved particles, which are
characterised by large values of \ptvecmiss and values of $\HTmiss /
\ETmiss$ close to unity.

Significant jet activity and \ptvecmiss in the event is ensured by
requiring $\scalht > 200\GeV$ and $\HTmiss > 130\GeV$,
respectively. These requirements complete the common preselection
criteria, summarised in Table~\ref{tab:selections}, used to define a
sample of all-jet events characterised by high jet activity and
appreciable \ptvecmiss.

\begin{table*}[tb]
  \topcaption{Summary of the event selection criteria and
    categorisation used to define the signal and control
    regions.}
  \label{tab:selections}
  \centering
  \cmsTable{\begin{tabular}{ ll }
    \hline
    \multicolumn{2}{l}{Common preselection}  \\[-1.5ex]                                                                                                     \\
    \ETmiss quality               & Filters related to beam and instrumental effects, and reconstruction failures                                     \\
    Lepton/photon vetoes          & $\pt > 10,\, 10,\, 25\GeV$ for isolated tracks, leptons, photons (respectively) and $\abs{\eta} < 2.5$            \\
    Jet $\mathrm{j}_i$ acceptance & Consider each jet $\mathrm{j}_i$ that satisfies $\pt^{\mathrm{j}_i} > 40\GeV$ and $\abs{\eta^{\mathrm{j_1}}} < 3$ \\
    Jet $\mathrm{j_1}$ acceptance & $\pt^{\mathrm{j_1}} > 100\GeV$ and $\abs{\eta^{\mathrm{j_1}}} < 2.5$                                              \\
    Jet $\mathrm{j_2}$ acceptance &
    $\pt^{\mathrm{j_2}} < 40\GeV$ (monojet),
    $40 < \pt^{\mathrm{j_2}} < 100\GeV$ (asymmetric),
    $\pt^{\mathrm{j_2}} > 100\GeV$ (symmetric)                                                                                                        \\
    Forward jet veto              & Veto events containing a jet satisfying $\pt > 40\GeV$ and $\abs{\eta} > 3$                                         \\
    Jets below threshold          & $\HTmiss / \ETmiss < 1.25$                                                                                        \\
    Energy sums                   & $\scalht > 200\GeV$ and $\HTmiss > 130\GeV$                                                                       \\
    \hline
    \multicolumn{2}{l}{Event categorisation}  \\[-1.5ex]                                                                                                    \\
    \njet                         & 1 (monojet), 2, 3, 4, $\geq$5 (asymmetric), 2, 3, 4, $\geq$5 (symmetric)                                          \\
    \nb                           & 0, 1, 2, $\geq$3 ($\nb \leq \njet$)                                                                               \\
    \scalht(\GeVns{})                 & 200, 250, 300, 350, 400, 500, 600, $>800\GeV$ (bins can be dropped/merged \vs \njet, Table~\ref{tab:binning})     \\
    \hline
    {Signal region (SR)}      & Preselection +                                                                                                    \\[1ex]
    QCD multijet rejection \quad  & $\alphat > 0.65$, 0.60, 0.55, 0.53, 0.52, 0.52, 0.52 (mapped to \scalht bins in range $200 < \scalht < 800\GeV$)  \\
    QCD multijet rejection        & $\bdphi > 0.5$ ($\njet \geq 2$) or $\bdphimod > 0.5$ ($\njet = 1$)                                               \\[0.5ex]
    \hline
    {Control regions (CR)}    & Preselection +                                                                                                    \\[1ex]
    Multijet-enriched             & SR + $\HTmiss/\ETmiss > 1.25$ (inverted)                                                                          \\
    \gj                           &
    1$\gamma$ with $\pt > 200\GeV$, $\abs{\eta} < 1.45$,
    $\Delta R(\gamma,\mathrm{j}_i) > 1.0$,
    $\scalht > 400\GeV$, same \alphat req. as SR                                                                                                      \\[0.5ex]
    \mj                           &
    1$\mu$ with $\pt > 30\GeV$, $\abs{\eta} < 2.1$,
    $I^{\mu}_\text{rel} < 0.1$,
    $\Delta R(\mu,\mathrm{j}_i) > 0.5$,
    $30 < m_\mathrm{T}(\ptvec^\mu,\ptvecmiss) < 125\GeV$                                                                                                \\[0.5ex]
    \mmjpm                        &
    2$\mu$ with $\pt > 30\GeV$, $\abs{\eta} < 2.1$,
    $I^{\mu}_\text{rel} < 0.1$,
    $\Delta R(\mu_{1,2},\mathrm{j}_i) > 0.5$,
    $ \abs{m_{\mu\mu} - m_\text{Z}} < 25\GeV$                                                                                                         \\[0.5ex]
    \hline
  \end{tabular}}
\end{table*}

\subsection{Event categorisation}
\label{sec:categorisation}

Events selected by the common preselection criteria are categorised
according to \njet, the number of b-tagged jets \nb, and \scalht. Nine
categories in \njet are employed: the monojet topology ($\njet = 1$)
and four \njet bins (2, 3, 4, $\geq$5) for each of the asymmetric and
symmetric topologies. Events are also categorised by \nb (0, 1, 2,
$\geq$3), where \nb is bounded from above by \njet, resulting in 32
categories in terms of both \njet and \nb. For each (\njet, \nb)
category, events are binned according to \scalht: four 50\GeV bins at
low jet activity in the range $200 < \scalht < 400\GeV$, two 100\GeV
bins in the range $400 < \scalht < 600\GeV$, one bin covering the
region $600 < \scalht < 800\GeV$, and a final open bin for $\scalht >
800\GeV$. These categorisations are summarised in
Table~\ref{tab:selections}. The \scalht binning scheme is adapted
independently per (\njet, \nb) category by removing or merging bins to
satisfy a threshold on the minimum number of data events in the
control regions, which are used to estimate SM backgrounds, provide
checks, and validate assumptions within the methods. The lower bounds
of the first and final (open) bins in \scalht are summarised in
Table~\ref{tab:binning}. In summary, the search employs a
categorisation scheme for events that results in 191 bins, defined in
terms of \njet, \nb, and \scalht.

\newcommand{\dash}{\multicolumn{1}{c}{\NA}}
\begin{table}[tb]
  \topcaption{Summary of the lower bounds of the first and final bins
    in \scalht [\GeVns{}] (the latter in parentheses) as a function of
    \njet and \nb.}
  \label{tab:binning}
  \centering
  {\begin{tabular}{ lrrrr }
    \hline
    $\njet \backslash\, \nb$ & 0         & 1         & 2         & $\geq$3                       \\
    \hline
    \multicolumn{5}{l}{Monojet}                                                              \\
    1                        & 200 (600) & 200 (500) & \dash     & \dash                         \\[1.5ex]
    \multicolumn{5}{l}{Asymmetric}                                                           \\
    2                        & 200 (600) & 200 (500) & 200 (400) & \dash                         \\
    3                        & 200 (600) & 200 (600) & 200 (500) & 200 (300)                     \\
    4                        & 200 (600) & 200 (600) & 200 (600) & 250 (400)                     \\
    $\geq$5                  & 250 (600) & 250 (600) & 250 (600) & 300 (500)                     \\[1.5ex]
    \multicolumn{5}{l}{Symmetric}                                                            \\
    2                        & 200 (800) & 200 (800) & 200 (600) & \dash                         \\
    3                        & 200 (800) & 250 (800) & 250 (800) & \phantom{0}\NA\phantom{0} (250) \\
    4                        & 300 (800) & 300 (800) & 300 (800) & 300 (800)                     \\
    $\geq$5                  & 350 (800) & 350 (800) & 350 (800) & 350 (800)                     \\
    \hline
  \end{tabular}}
\end{table}

\subsection{Signal region}
\label{sec:signal_region}

For events satisfying the common preselection criteria described
above, the multijet background dominates over all other SM
backgrounds. Several variables are employed to reduce the multijet
contribution to a low level with respect to other SM backgrounds.

The dimensionless kinematic variable \alphat~\cite{Randall:2008rw,
  RA1Paper}, defined in Eq.~(\ref{eq:alphat}) below, is used to
provide discrimination against multijet events that do not contain
significant \ptvecmiss or that contain large \ptvecmiss only because
of \pt mismeasurements, while retaining sensitivity to new-physics
events with significant \ptvecmiss. The \alphat variable depends
solely on the transverse component of jet four-momenta
and is intrinsically robust against the presence of jet energy
mismeasurements in multijet systems. For events containing only two
jets, \alphat is defined as $\alphat = \ET^{\mathrm{j}_2}/M_\mathrm{T}$,
where $\ET= E\sin\theta$,
where $E$ is the energy of the jet and $\theta$ is its polar angle
with respect to the beam axis, $\ET^{\mathrm{j}_2}$ is the transverse
energy of the jet with smaller \ET, and $M_\mathrm{T}$ is the transverse
mass of the dijet system, defined as:
\begin{equation}
  \label{eq:mt}
  M_\mathrm{T} = \sqrt{ \Bigl( \sum_{i=1,2} \ET^{\mathrm{j}_i}
    \Bigr)^2 - \Bigl( \sum_{i=1,2} p_x^{\mathrm{j}_i} \Bigr)^2 - \Bigl(
      \sum_{i=1,2} p_y^{\mathrm{j}_i} \Bigr)^2}\, ,
\end{equation}
where $\ET^{\mathrm{j}_i}$, $p_x^{\mathrm{j}_i}$, and
$p_y^{\mathrm{j}_i}$ are, respectively, the transverse energy, and the
$x$ and $y$ components of the transverse momentum of jet
$\mathrm{j}_i$.
For a perfectly measured dijet event with $\ET^{\mathrm{j_1}} =
\ET^{\mathrm{j_2}}$ and back-to-back jets ($\Delta\phi = \pi$), and in
the limit in which the momentum of each jet is large compared with its
mass, the value of \alphat is 0.5. For an imbalance in the \ET of
back-to-back jets, \alphat is reduced to a value $<$0.5, which gives
the variable its intrinsic robustness. Values significantly greater
than 0.5 are observed when the two jets are not back-to-back and
recoil against \ptvecmiss from weakly interacting particles that
escape the detector.

The definition of the \alphat variable can be generalised for events
with more than two jets~\cite{RA1Paper}. The mass scale for any
process is characterised through the scalar sum of the jet transverse
energies, defined as $\scalst = \sum_{i=1}^{N_\text{jet}}
\ET^{\mathrm{j}_i}$, where $N_\text{jet}$ is the number of jets with
\ET above a predefined threshold. (The definition of \scalst should be
contrasted with that of \scalht, the scalar \pt sum of the jets.)
For events with three or more jets, a pseudo-dijet system is formed by
combining the jets in the event into two pseudo-jets. The \scalst for
each of the two pseudo-jets is given by the scalar \ET sum of its
contributing jets. The combination chosen is the one that minimises
\dst, defined as the difference between these sums for the two
pseudo-jets.  This clustering criterion assumes a balanced-event
hypothesis, which provides strong separation between SM multijet
events and events with genuine \ptvecmiss. The \alphat definition can
be generalised to:
\begin{equation}
  \label{eq:alphat}
  \alphat = \frac{1}{2} \frac{\scalst -
    \dst}{\sqrt{(\scalst)^2 - (\HTmiss)^2}}.
\end{equation}

When jet energies are mismeasured, or there are neutrinos from
heavy-flavour quark decays, the magnitudes of \HTmiss and \dst are
highly correlated. This correlation is much weaker for
$R$-parity-conserving SUSY events, where each of the two decay chains
produces an undetected LSP.

Multijet events populate the region $\alphat< 0.5$ and the $\alphat$
distribution is characterised by a sharp edge at 0.5, beyond which the
multijet event yield falls by several orders of magnitude. Multijet
events with extremely rare but large stochastic fluctuations in the
calorimetric measurements of jet energies can lead to values of
\alphat slightly above 0.5. The edge at 0.5 sharpens with increasing
\scalht for events containing at least three jets, primarily due to a
corresponding increase in the average jet energy and consequently a
(relative) improvement in the jet energy resolution.

For events containing at least two jets, thresholds on the minimum
allowed \alphat values are applied independent of \njet and \nb but
dependent on \scalht, for events that satisfy $200 < \scalht <
800\GeV$. The \alphat thresholds vary between 0.65 to 0.52 for,
respectively, the regions $200 < \scalht < 250\GeV$ and $400 < \scalht
< 800\GeV$. No requirement on \alphat is made for the region $\scalht
> 800\GeV$. The thresholds employed are summarised in
Table~\ref{tab:selections}. The \alphat thresholds are motivated both
by the trigger conditions used to record the candidate signal events,
described below, and by simulation-based studies and estimates of the
multijet background derived from data.

An additional variable is based on the minimum azimuthal separation
between a jet and the negative vector \ptvec sum derived from all
other jets in the event~\cite{RA1Paper},
\begin{equation}
  \bdphi = \min_{\,\forall\, \mathrm{j}_k\,\in\, [1,\njet]}
  \Delta\phi \Bigl( \ptvec^{\,\mathrm{j}_k}, \,
    -\hspace{-0.5em}\sum_{\substack{\mathrm{j}_i= 1 \\ \mathrm{j}_i \ne \mathrm{j}_k}}^{\njet}
    \ptvec^{\,\mathrm{j}_i} \Bigr).
  \label{eq:bdphi}
\end{equation}

This variable discriminates between final states with genuine
\ptvecmiss, \eg from the leptonic decay of the W boson, and energetic
multijet events that have significant \ptvecmiss through jet energy
mismeasurements or through the production of neutrinos, collinear with
the axis of a jet, from semileptonic heavy-flavour decays. Multijet
events populate the region $\bdphi < 0.5$, with the multijet
distribution peaking at a value of zero and falling approximately
exponentially over five orders of magnitude to a single-event level at
a value of $\bdphi \approx 0.5$, which is close to the distance
parameter value of 0.4 used by the anti-\kt jet clustering
algorithm. Events with a genuine source of \ptvecmiss exhibit a long
tail in \bdphi with values as large as $\pi$.

\begin{figure}[!htb]
  \centering
    \includegraphics[width=0.49\textwidth]{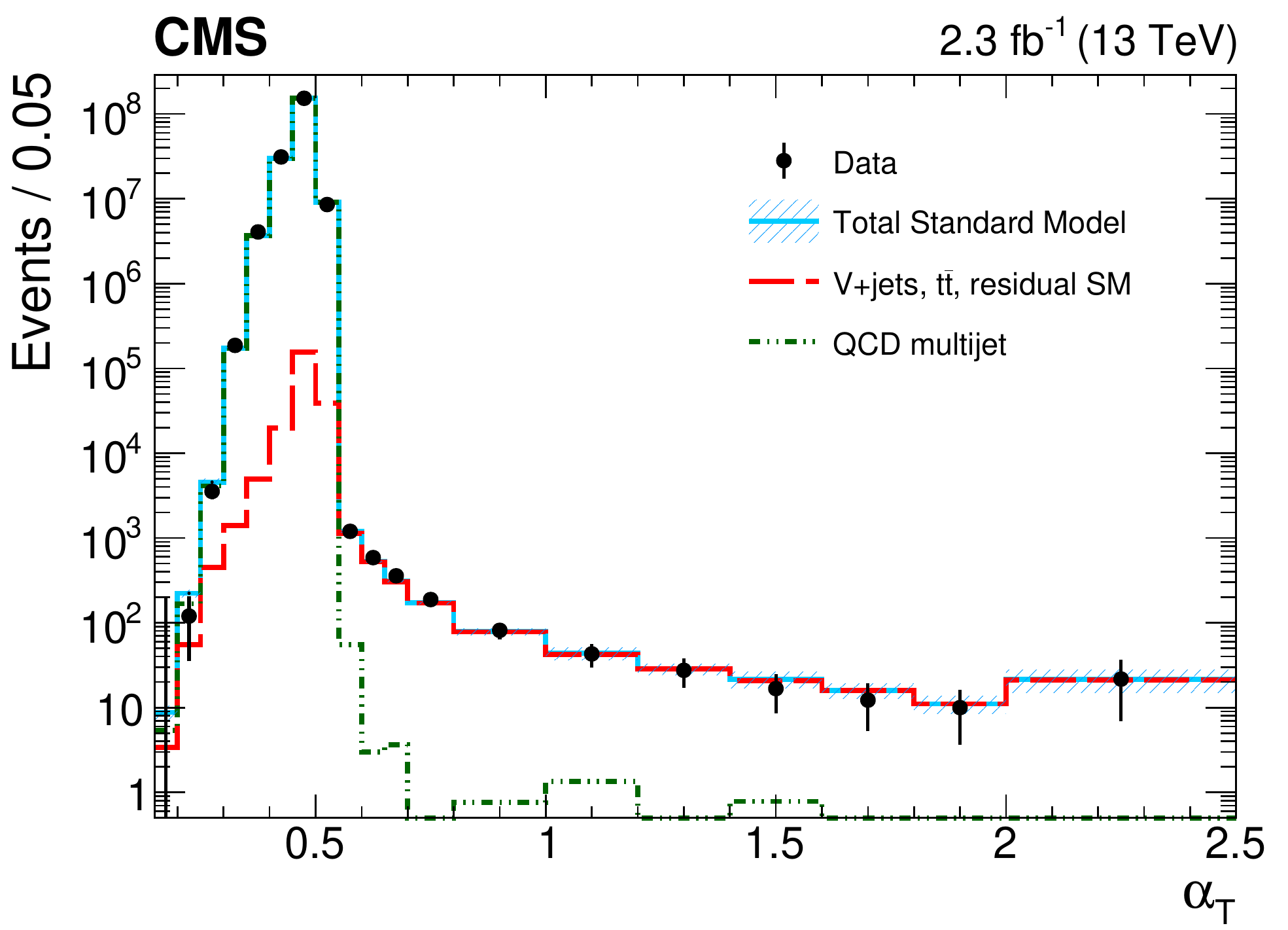} \,
    \includegraphics[width=0.49\textwidth]{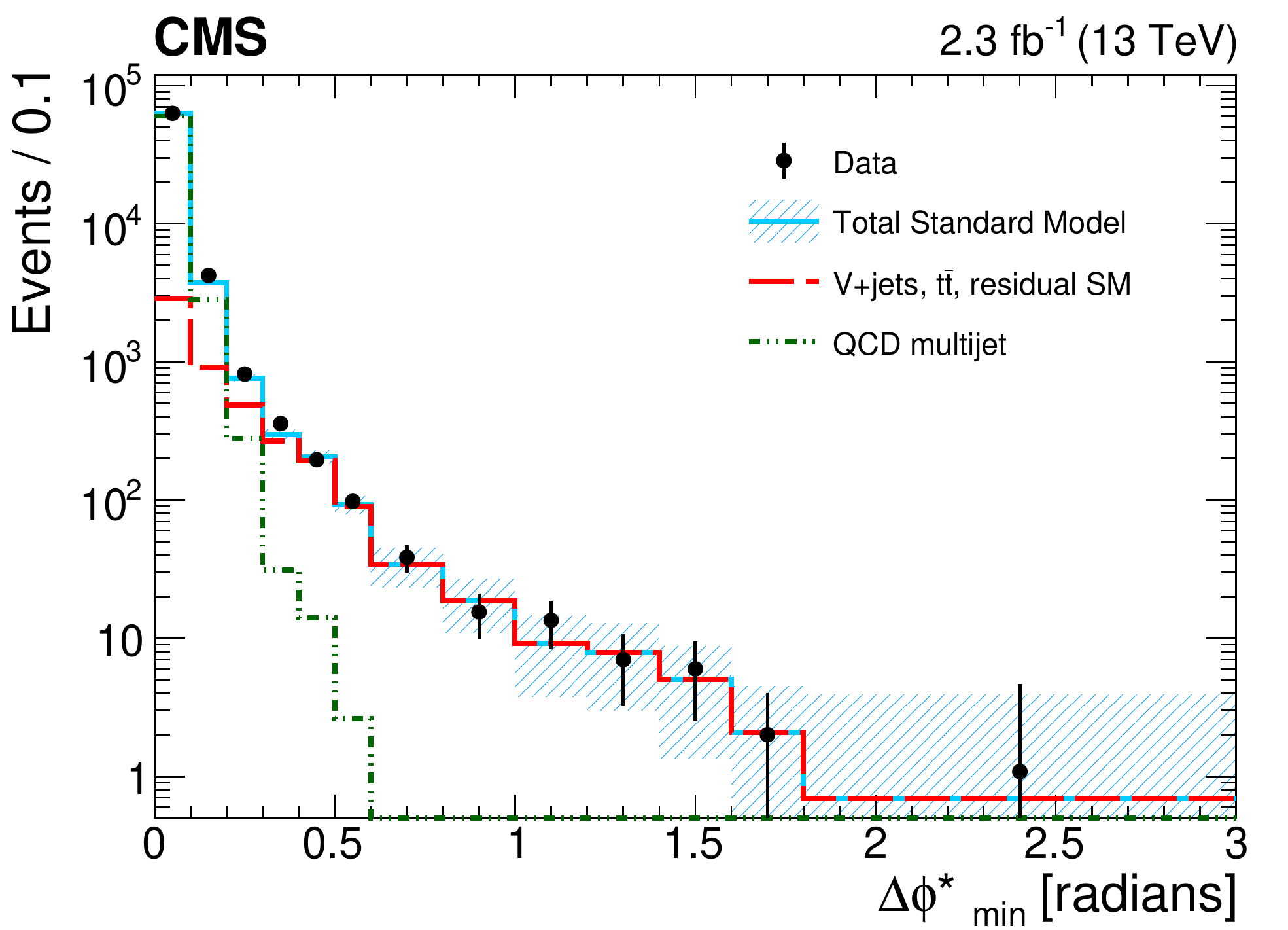} \\
  \caption{
    The (\cmsLeft) \alphat and (\cmsRight) \bdphi distributions observed in
    data for events that satisfy the selection criteria defined in the
    text. The statistical uncertainties for the multijet and SM
    expectations are represented by the hatched areas (visible only
    for statistically limited bins). The final bin of each
    distribution contains the overflow events.  }
  \label{fig:alphat-bdphi}
\end{figure}

A requirement of $\bdphi > 0.5$ is sufficient to effectively suppress
the multijet background to a low level while maintaining high
efficiency for new-physics signatures. The combined rejection power of
the \alphat and \bdphi requirements for the region $200 < \scalht <
800\GeV$ is sufficient to suppress multijet events to the few percent
level (and always $<$10\%) with respect to all other SM backgrounds in
all \scalht bins for all event categories of the signal region. For
the region $\scalht > 800\GeV$, a similar control of the multijet
background is achieved solely with the $\bdphi > 0.5$ requirement.

Figure~\ref{fig:alphat-bdphi} shows the distributions of \alphat and
\bdphi observed in data for events that satisfy the full set of
selection criteria used to define the signal region, summarised in
Table~\ref{tab:selections}, as well as the following
modifications. The \alphat and \bdphi distributions are constructed
from events that satisfy $\njet \geq 2$, $\pt^{\mathrm{j_2}} >
100\GeV$, and, respectively, $\scalht > 300$ or $800\GeV$. In the case
of Fig.~\ref{fig:alphat-bdphi} (\cmsLeft), the events with $\alphat$
values greater than 0.55 must fulfill the full set of signal region
criteria, including the $\bdphi > 0.5$ requirement, while the events
that satisfy $\alphat < 0.55$ are subject to the looser set of common
preselection criteria defined in Table~\ref{tab:selections},
excluding the $\HTmiss > 130\GeV$ requirement. Hence,
Fig.~\ref{fig:alphat-bdphi} (\cmsLeft) demonstrates the combined
performance of several variables that are employed to suppress
multijet events. For both distributions, the events are recorded with
a set of inclusive trigger conditions that are independent of the
\alphat and \bdphi variables. The distributions for the QCD multijet
background are determined from simulation while all other SM
backgrounds (vector boson production in association with jets, \ttbar,
and other residual contributions from rare SM processes) are estimated
using a \mj data control sample, described in
Section~\ref{sec:ewk_background}. The contribution from multijet
events is observed to fall by more than five orders of magnitude for
both variables.

The \alphat and \bdphi requirements described above, in conjunction
with the common preselection requirements $\HTmiss > 130\GeV$ and
$\HTmiss / \ETmiss < 1.25$, provide strong rejection power against the
multijet background for events that satisfy $\njet \geq 2$. For
monojet events, a modification to the \bdphi variable, which considers
soft jets with $\pt > 25\GeV$ ($\bdphimod$), is utilised. No \alphat
requirement is imposed, and $\bdphimod > 0.5$ is sufficient to
suppress contributions from multijet events to a negligible level.
The aforementioned requirements complete the event selection criteria
for the signal region.

Finally, \bdphimod is also used as a control variable for events that
satisfy $\njet \geq 2$ to identify multijet contributions arising from
instrumental effects, such as inefficient detector elements or
detector noise. The axis of any jet that satisfies $\bdphimod < 0.5$
is used to identify localised behaviour in the ($\eta$, $\phi$) plane,
which may be indicative of instrumental defects. No significant
anomalies are observed in the sample of candidate signal events
following the application of the dedicated vetoes described in
Section~\ref{sec:preselection}.

{\tolerance=800
Multiple trigger conditions are employed in combination to record 
candidate signal events. A set of trigger conditions utilise
calculations of both \scalht and \alphat to record events with two or
more jets. An event is recorded if it satisfies any of the following
pairs of (\scalht [\GeVns{}], \alphat) thresholds, (200, 0.57), (250,
0.55), (300, 0.53), (350, 0.52), or (400, 0.51), as well as a
requirement on the mean value of the two highest \pt jets,
$\langle\pt^{\mathrm{j_1}} + \pt^{\mathrm{j_2}}\rangle >
90\GeV$. These requirements are collectively labelled as the
``\scalht--\alphat'' triggers. In addition, candidate signal events
with one or more jets are also recorded if they satisfy the
requirements $\HTmiss > 90\GeV$ and $\ETmiss > 90\GeV$. Finally, for
events that satisfy $\scalht > 800\GeV$, an additional trigger
condition, defined by $\scalht > 800\GeV$, is employed in addition to
the \scalht--\alphat trigger requirements to record events
characterised by high activity in the calorimeters with high
efficiency. The trigger-level jet energies are corrected to account
for energy scale and pileup effects. The aforementioned triggers are
employed in combination to provide efficiencies at or near 100\% for
all bins in the signal region.
\par}

\subsection{Using \texorpdfstring{\HTmiss}{HTmiss} templates}
\label{sec:mht_templates}

Following the event selection criteria described above, which provide
a sample of candidate signal events with a negligible contribution
from multijet events, further discriminating power is required to
separate new-physics signatures from the remaining SM backgrounds,
which are dominated by the production of \ttbar or \wlj and \znunuj
events. As discussed in Sec.~\ref{sec:introduction}, the production of
coloured SUSY particles with decays to a weakly interacting LSP
typically gives rise to a final state with multiple jets and large
\ptvecmiss. The search therefore exploits the \HTmiss variable as an
additional discriminant between new-physics and SM processes.

The search relies directly on simulation to determine a template for
each (\njet, \nb, \scalht) bin that describes the expected
distribution of events as a function of \HTmiss. These templates are
used by the likelihood function as a model for the data, details of
which can be found in Section~\ref{sec:result}. The templates are
extensively validated against data in multiple control regions, and
these studies are used to establish the uncertainty in the simulation
modelling of the \HTmiss variable. The effects of theoretical and
experimental uncertainties on the \HTmiss distributions are also
studied. Further details can be found in
Section~\ref{sec:ewk_background}.

\subsection{Control regions}
\label{sec:control_regions}

Four control regions in data are employed to estimate the background
contributions from SM processes. The event selection criteria used to
define the control regions comprise the common preselection
requirements and additional sample-specific requirements, as
summarised in Table~\ref{tab:selections}. The first control region
comprises a multijet-enriched sample of events, and is defined by the
signal region selection criteria and the inverted requirement $\HTmiss
/ \ETmiss > 1.25$. The events are recorded with the signal triggers
described above, and the sample is used to estimate the multijet
background in the signal region. Three additional control regions,
defined by inverting one of the photon or lepton vetoes to select
samples of \gj, \mj, or \mmj events, are used to estimate the
background contributions from SM processes with final states
containing genuine \ptvecmiss, which are primarily \ttbar, \wlj, and
\znunuj.

Additional kinematic requirements are employed to ensure the control
samples are enriched in the same SM processes that contribute to
background events in the signal region, and are depleted in
contributions from multijet production or a wide variety of SUSY
models (\ie so-called signal contamination).  The samples are defined,
and their events are identically categorised and binned, such that the
kinematic properties of events in the control regions and the
candidate signal events resemble as closely as possible one another
once the photon, muon, or dimuon system is ignored in the calculation
of quantities such as \scalht and \HTmiss. The selections are
summarised in Table~\ref{tab:selections} and described below.

The \gj event sample is defined by the common preselection
requirements, but the photon veto is inverted and each event is
required to contain a single isolated photon, as defined in
Section~\ref{sec:event_reconstruction}, that satisfies $\pt > 200\GeV$
and $\abs{\eta} < 1.45$ and is well separated from each jet
$\mathrm{j}_i$ in the event according to $\Delta
R(\gamma,\mathrm{j}_i) > 1.0$.
In addition, events must satisfy $\scalht > 400\GeV$, as well as the
same \scalht-dependent \alphat requirements used to define the signal
region. The events are recorded using a single-photon trigger
condition and the selection criteria result in a trigger efficiency
of~$\gtrsim$99\%.

The \mj event sample is defined by the common preselection
requirements, but the muon veto is inverted and each event is required
to contain a single isolated muon, as defined in
Section~\ref{sec:event_reconstruction}, that satisfies $\pt > 30\gev$
and $\abs{\eta} < 2.1$ and is well separated from each jet
$\mathrm{j}_i$ in the event according to $\Delta R(\mu,\mathrm{j}_i) >
0.5$.
The transverse mass formed by the muon \pt and \ptvecmiss system must
satisfy $30 < m_\mathrm{T} < 125\GeV$ to select a sample of events rich
in W bosons, produced promptly or from the decay of top quarks. The
\mmj sample uses a similar set of selection criteria as the \mj
sample, but specifically requires two oppositely charged isolated
muons that both satisfy $\pt > 30\gev$ and $\abs{\eta} < 2.1$ and are
well separated from the jets in the event ($\Delta
R(\mu_{1,2},\mathrm{j}_i) > 0.5$). The muons are also required to have
a dilepton invariant mass within a $\pm 25\GeV$ window around the
nominal mass of the Z boson~\cite{1674-1137-38-9-090001}. For both the
muon and dimuon samples, no requirement is made on \alphat in order to
increase the statistical precision of the predictions from these
samples. Both the \mj and \mmj samples are recorded using a trigger
that requires an isolated muon. The selection criteria of the \mj and
\mmj event samples are chosen so that the trigger is maximally
efficient, with values of $\sim$90\% and $\sim$99\%, respectively.

\section{Estimation of backgrounds}
\label{sec:backgrounds}

\subsection{Multijet background}
\label{sec:qcd_background}

The signal region is defined in a manner that suppresses the expected
contribution from multijet production to a low level with respect to
the total expected background from other SM processes for all signal
region bins. This is achieved primarily through the application of
very tight requirements on the variables \alphat and \dphi, as
described in Section~\ref{sec:signal_region}, as well as the
requirement $\mhtmet < 1.25$. In this section, we discuss these
requirements further, and present the estimate of the multijet
background.

The contamination from multijet events in the signal region is
estimated using a multijet-enriched data sideband to the signal
region, defined by the (inverted) requirement $\mhtmet > 1.25$. The
observed counts in data, categorised according to \njet and \scalht,
are corrected to account for contamination from nonmultijet SM
processes, and the corrected counts $\mathcal{N}^\text{data}(\njet,
\scalht)$ are assumed to arise solely from QCD multijet
production. The nonmultijet processes, which comprise vector boson
and \ttbar production and residual contributions from other SM
processes, are estimated using the \mj control region, as
described in Section~\ref{sec:ewk_background}.

Independent ratios $\mathcal{R}^\mathrm{QCD}(\njet, \scalht)$ of the
number of multijet events that satisfy the requirement $\mhtmet <
1.25$ to the number that fail this requirement are determined from
simulation for events categorised according to \njet and \scalht, and
inclusively with respect to \nb and \HTmiss. The product of each ratio
$\mathcal{R}^\mathrm{QCD}(\njet, \scalht)$ and the corresponding
corrected data count $\mathcal{N}^\text{data}(\njet, \scalht)$
provides the estimate of the multijet background $\mathcal{P}(\njet,
\scalht)$. The estimates as a function of \njet, \scalht, \nb, and
\HTmiss of the signal region are assumed to factorise as follows:
\begin{equation}
  \label{eq:qcd1}
  \mathcal{P}( \njet, \scalht )  =
  \mathcal{N}^\text{data}( \njet, \scalht )\;
  \mathcal{R}^\mathrm{QCD}( \njet, \scalht ),
\end{equation}
\begin{equation}
  \label{eq:qcd2}
  \mathcal{P}( \njet, \scalht, \nb, \HTmiss ) = 
  \mathcal{P}( \njet, \scalht )\;
  \mathcal{K}_{\njet, \scalht}( \nb, \HTmiss ), 
\end{equation}
where $\mathcal{K}_{\njet, \scalht}( \nb, \HTmiss )$ are multiplier
terms that provide the estimated distribution of events as a function
of \nb and \HTmiss while preserving the normalisation $\mathcal{P}(
\njet, \scalht )$.

\begin{figure*}[!thb]
  \centering
    \includegraphics[width=\cmsFigWidthTwo]{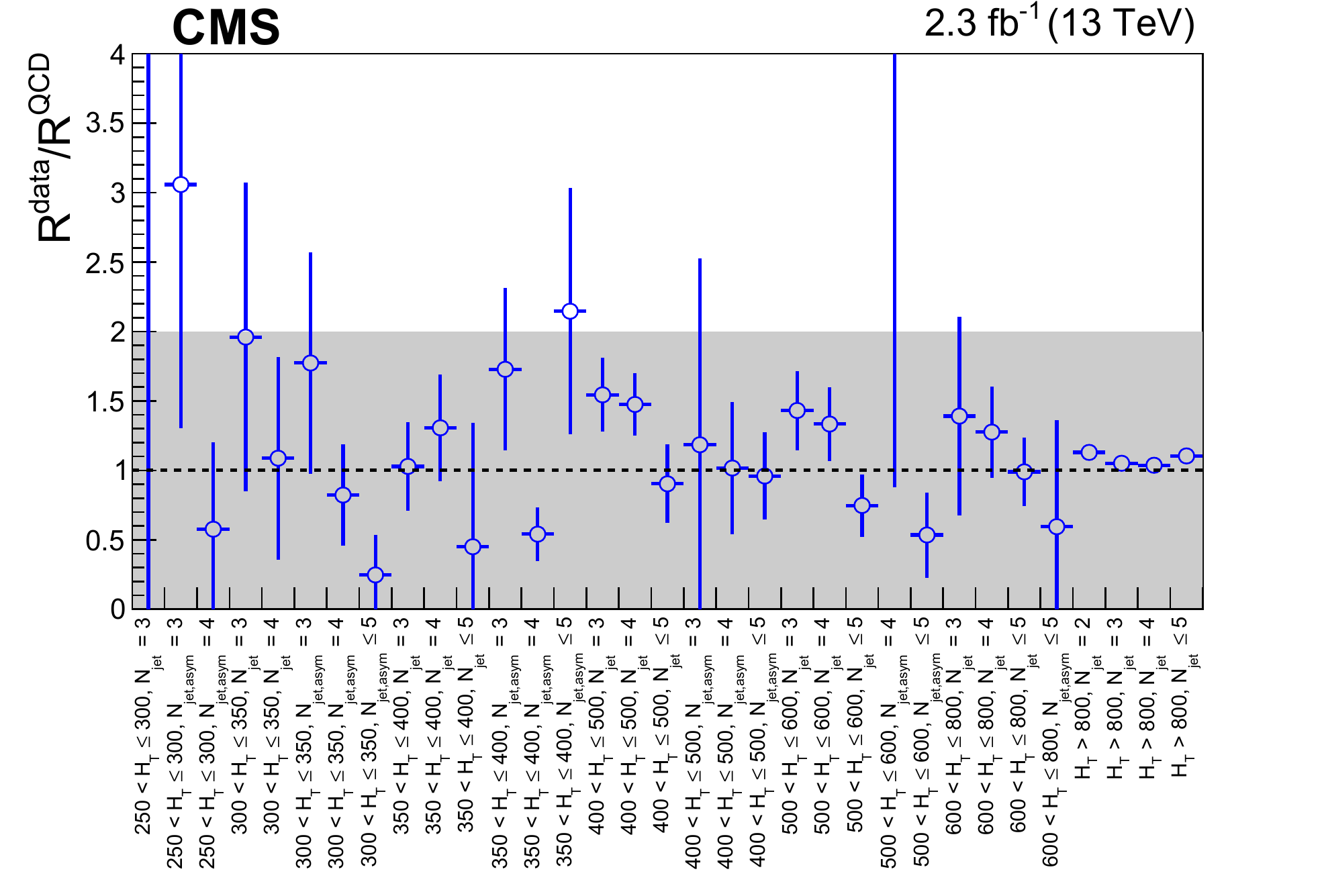} \\
  \caption{Validation of the ratio $\mathcal{R}^\mathrm{QCD}$ determined
    from simulation in bins of \njet and \scalht [\GeVns{}] by comparing
    with an equivalent ratio $\mathcal{R}^\text{data}$ constructed
    from data in a multijet-enriched sideband to the signal region. A
    value of unity is expected for the double ratio
    $\mathcal{R}^\text{data} / \mathcal{R}^\mathrm{QCD}$, and the grey
    shaded band represents the assumed systematic uncertainty of 100\%
    in $\mathcal{R}^\mathrm{QCD}$.  }
  \label{fig:qcd}
\end{figure*}

The use of simulation to determine $\mathcal{R}^\mathrm{QCD}(\njet,
\scalht)$ is validated using a multijet-enriched data sideband defined
by $\bdphi < 0.5$. Each ratio $\mathcal{R}^\text{data}(\njet,
\scalht)$ is constructed from data counts, corrected to account for
contributions from nonmultijet processes, and compared with the
corresponding ratio $\mathcal{R}^\mathrm{QCD}(\njet, \scalht)$,
determined from simulation, through the double ratio
$\mathcal{R}^\text{data}/\mathcal{R}^\mathrm{QCD}$, as shown in
Fig.~\ref{fig:qcd}. The double ratios are statistically compatible
with unity across the full phase space of the signal region, including
the bins at high \scalht, which exhibit the highest statistical
precision. In addition to statistical uncertainties as large as
$\sim$100\%, a systematic uncertainty of 100\% in
$\mathcal{R}^\mathrm{QCD}$ is assumed to adequately cover the observed
level of agreement for the full signal region phase space. 

The distribution of multijet events as a function of \nb and \HTmiss,
$\mathcal{K}_{\njet, \scalht}( \nb, \HTmiss )$, is assumed to be
identical to the distribution expected for the nonmultijet
backgrounds. This final assumption is based on studies in simulation
and is a valid approximation given the magnitude of this background
contribution, as well as the magnitude of the statistical and
systematic uncertainties in the ratios
$\mathcal{R}^\mathrm{QCD}(\njet, \scalht)$, as described above.
 
\subsection{Backgrounds with genuine \texorpdfstring{\ETmiss}{MET}}
\label{sec:ewk_background}

Following the suppression of multijet events through the use of the
\alphat and \bdphi variables, the dominant nonmultijet backgrounds
involve SM processes that produce high-\pt neutrinos in the final
state. In events with few jets or few b quark jets, the associated
production of W or Z bosons and jets, with the decays $\PW^\pm
\to \ell\nu$ ($\ell=\Pe$, $\Pgm$, $\Pgt$) or \znunu,
dominate the background counts. For W boson decays that yield an
electron or muon (possibly originating from leptonic $\Pgt$ decays),
the background contributions result from events containing an $\Pe$ or
$\Pgm$ that are not rejected by the lepton vetoes.
The veto of events containing at least one isolated track further
suppresses these backgrounds, including those from single-prong
$\tau$-lepton decays. At higher jet or b-quark jet multiplicities,
single top quark and \ttbar production, followed by semileptonic top
quark decay, also become an important source of background.

The method to estimate the nonmultijet backgrounds in the signal
region relies on the use of a transfer (\tf) factor determined from
simulation that is constructed per bin (in terms of \njet, \nb, and
\scalht) per control region. Each \tf factor is defined as the ratio
of the expected yields in the same (\njet, \nb, \scalht) bins of the
signal region $\mathcal{N}^\text{SR}_\text{MC}$ and one of the control
regions $\mathcal{N}^\text{CR}_\text{MC}$.  The \tf factors are used to
extrapolate from the event yields observed in each bin of a data
control sample $\mathcal{N}^\text{CR}_\text{data}$ to provide an
estimate for the background, integrated over \HTmiss, from a
particular SM process or processes in the corresponding bin of the
signal region $\mathcal{N}^\text{SR}_\text{data}$.  The superscript SR
or CR refers to, respectively, the process or processes being
estimated and one of the \mj, \mmj, and \gj control regions, described
in Section~\ref{sec:control_regions}. The subscript refers to whether
the counts are obtained from data, simulation (``MC''), or an estimate
(``pred'').

The method aims to minimise the effects of simulation mismodelling, as
many systematic biases in the simulation are expected to largely
cancel in the \tf factors, given that the events in any given (\njet,
\nb, \scalht) bin of the control regions closely mirror those in the
corresponding bin in the signal region in terms of the event energy
scale, topology, and kinematics. In short, minimal extrapolations are
made. Uncertainties in the \tf factors are determined from data, as
described below.

Three independent estimates of the irreducible background of \znunu +
jets events are determined from the \gj, \mmj, and \mj data control
samples. The \gj and \zmumu + jets processes have similar kinematic
properties when the photon or muons are ignored~\cite{Bern:2011pa},
albeit different acceptances. In addition, the \gj process has a
larger production cross section than \znunu + jets events. The \mj
data sample is used to provide an estimate for both the \znunu\ + jets
background, as well as the other dominant SM processes, \ttbar and W
boson production (labelled collectively as $\PW/\ttbar$). Residual
contributions from all other SM relevant processes, such as single top
quark, diboson, and Drell--Yan production, are also included as part
of the $\PW/\ttbar$ estimate from the \mj sample. The definition of
the various \tf factors used in the search are given below:
\begin{align}
  \mathcal{N}^\text{\PW/\ttbar}_\text{pred} \, & = \,
  \tf^\text{\PW/\ttbar}_{\mj} \;
  \mathcal{N}^\text{\mj}_\text{data}, &
  \tf^\text{\PW/\ttbar}_{\mj} \, & = \,
  \bigg(
  \frac{\,\mathcal{N}^{\PW/\ttbar}_\text{MC}\,}
  {\mathcal{N}^\text{\mj}_\text{MC}}
  \bigg); \\
  \mathcal{N}^{\znunu}_\text{pred} \, & = \,
  \tf^{\text{\znunu}}_{\mj} \;
  \mathcal{N}^\text{\mj}_\text{data}, &
  \tf^{\text{\znunu}}_{\mj} \, & = \,
  \bigg(
  \frac{\,\mathcal{N}^\text{\znunu}_\text{MC}\,}
  {\mathcal{N}^\text{\mj}_\text{MC}}
  \bigg); \\
  \mathcal{N}^\text{\znunu}_\text{pred} \, & = \,
  \tf^{\znunu}_{\mmj}  \;
  \mathcal{N}^\text{\mmj}_\text{data}, &
  \tf^{\znunu}_{\mmj}  \, & = \,
  \bigg(
  \frac{\,\mathcal{N}^\text{\znunu}_\text{MC}\,}
  {\mathcal{N}^\text{\mmj}_\text{MC}}
  \bigg); \\
  \mathcal{N}^\text{\znunu}_\text{pred} \, & = \,
  \tf^{\znunu}_{\gj}  \;
  \mathcal{N}^\text{\gj}_\text{data}, &
  \tf^{\znunu}_{\gj}  \, & = \,
  \bigg(
  \frac{\,\mathcal{N}^\text{\znunu}_\text{MC}\,}
  {\mathcal{N}^\text{\gj}_\text{MC}}
  \bigg).
\end{align}

The likelihood function, described in Section~\ref{sec:result},
encodes the estimate via the \tf factors of the $\PW/\ttbar$ background, as
well as the three independent estimates of the \znunu background,
which are considered simultaneously.

Several sources of uncertainty in the \tf factors are evaluated.  The most
relevant effects are discussed below, and generally fall into one of
two categories. The first category concerns uncertainties in the
``scale factor'' corrections applied to simulation, which are
determined using inclusive data samples that are defined by loose
selection criteria, to account for the mismodelling of theoretical and
experimental parameters. The second category concerns ``closure
tests'' in data that probe various aspects of the accuracy of the
simulation to model correctly the \tf factors in the phase space of this
search.

{\tolerance=500
The uncertainties in the \tf factors are studied for variations in
scale factors related to the jet energy scale (that result in
uncertainties in the \tf factors as large as $\sim$15\%), the
efficiency and misidentification probability of b quark jets (up to
5\%), and the efficiency to identify well-reconstructed, isolated
leptons (up to $\sim$5\%). A 5\% uncertainty in the total inelastic
cross section, $\sigma_\text{in} = 69.0 \pm
3.5\unit{mb}$~\cite{Aaboud:2016mmw}, is assumed and propagated through
to the reweighting procedure to account for differences between the
simulated measured pileup, which results in changes of up to
$\sim$10\%. The modelling of the transverse momentum of top quarks
($\pt^\text{t}$) is evaluated by comparing the simulated and measured
\pt spectra of reconstructed top quarks in \ttbar
events~\cite{Khachatryan:2015oqa}.  Simulated events are reweighted
according to scale factors that decrease from a value of $\sim$1.2 to
$\sim$0.7, with uncertainties of $\sim$10--20\%, within the range
$\pt^\text{t} < 400\GeV$.
The systematic uncertainties in $\tf^\text{\PW/\ttbar}_{\mj}$ arising
from variations in the $\pt^\text{t}$ scale factors are typically
small ($\lesssim$5\%), due to the comparable phase space probed by the
signal and control regions, while larger uncertainties
($\lesssim$20\%) in $\tf^{\text{\znunu}}_{\mj}$ are observed due to
the potential for significant contamination from \ttbar when using
\wlj to predict \znunuj.
\par}

\newcommand{\phh}{\ensuremath{\phantom{1-}}}
\begin{table*}[h!]
  \topcaption{
    Systematic uncertainties (in percent) in the transfer (\tf) factors
    used in the method to estimate the SM backgrounds with genuine
    \ptvecmiss in the signal region. The quoted ranges provide
    representative values of the observed variations as a function of
    \njet and \scalht.
  }
  \label{tab:bkgd_systs}
  \centering
  \begin{tabular}{ lrrrr }
    \hline
    Systematic source            & \multicolumn{4}{c}{Uncertainty in \tf factor [\%]} \\
    \cline{2-5}
                                 & $\tf^{\PW/\ttbar\T}_{\mj\B}$
                                 & $\tf^{\znunu}_{\mj}$
                                 & $\tf^{\znunu}_{\mmj}$
                                 & $\tf^{\znunu}_{\gj}$                               \\
    \hline
    \multicolumn{5}{l}{Scale factors (applied to simulation):}                    \\[1ex]
    Jet energy scale             & $<$15    & $<$15   & $<$10   & $<$15               \\
    b tagging eff \& mistag rate & $<$5     & $<$5    & $<$2    & $<$2                \\
    Lepton identification        & 2--5    & 2--5   & 2--5   & \NA                \\
    Pileup                       & $<$10    & $<$6    & $<$4    & $<$3                \\
    Top quark \pt                & $<$5     & $<$20   & $<$4    & \NA                \\ [2ex]
    \multicolumn{5}{l}{Closure tests:}                                            \\[1ex]
    W/Z ratio                    & \NA     & 10--30 & \NA    & \NA                \\
    Z/$\gamma$ ratio             & \NA     & \NA    & \NA    & 10--30             \\
    W/\ttbar composition         & 10--100 & \NA    & \NA    & \NA                \\
    W polarisation               & 5--50   & 5--50  & \NA    & \NA                \\
    $\alphat\,/\,\bdphi$\B       & 5--80   & 5--80  & 50--80 & \NA                \\
    \hline
  \end{tabular}
\end{table*}

The aforementioned systematic uncertainties, resulting from variations
in scale factors, are summarised in Table~\ref{tab:bkgd_systs}, along
with representative magnitudes. These sources of uncertainty are each
assumed to originate from a unique underlying source and so the effect
of each source is varied assuming a fully correlated behaviour across
the full phase space of the signal and control regions.

The second category of uncertainty is determined from sets of closure
tests based on data control samples~\cite{RA1Paper2012}. Each set uses
the observed event counts in up to eight bins in \scalht for each of
the nine \njet event categories in one of the three independent data
control regions. These counts are used with the corresponding \tf
factors, determined from simulation, to obtain a prediction
$\mathcal{N}^\text{pred}(\njet, \scalht)$ of the observed yields
$\mathcal{N}^\text{obs}(\njet, \scalht)$ in another control sample
(or, in one case, \nb event category).

Each set of tests targets a specific (potential) source of bias in the
simulation modelling that may introduce an \njet- or \scalht-dependent
source of systematic bias in the \tf factors~\cite{RA1Paper2012}. Several
sets of tests are performed. The $\PZ/\gamma$ ratio determined from
simulation is tested against the same ratio measured using \zmumuj
events and the \gj sample.
The $\PW/\PZ$ ratio is also probed using the \mj and \mmj
samples, which directly tests the simulation modelling of vector
boson production, as well as the modelling of \ttbar contamination in
the \mj sample.
A further set probes the modelling of the relative composition between
\wlj and \ttbar events using \mj events containing exactly zero or one
more b-tagged jets, which represents a larger extrapolation in
relative composition than used in the search.  The effects of W
polarisation are probed by using \mj events with a positively charged
muon to predict those containing a negatively charged muon. Finally,
the accuracy of the modelling of the efficiencies of the \alphat and
\bdphi requirements are estimated using the \mj sample.

For each set of tests, the level of closure,
($\mathcal{N}^\text{obs} - \mathcal{N}^\text{pred}) /
\mathcal{N}^\text{obs}$, which considers only statistical
uncertainties, is inspected to ensure no statistically significant
biases are observed as a function of the nine \njet categories or the
eight \scalht bins. In the absence of such a bias, the level of
closure is recomputed by integrating over either all monojet and
asymmetric \njet categories, or the symmetric \njet categories. The
level of closure and its statistical uncertainty are combined in
quadrature to determine additional contributions to the uncertainties
in the \tf factors. These uncertainties are considered to be
fully correlated between the monojet and asymmetric \njet categories
or the symmetric \njet categories, and fully uncorrelated between
these two regions in \njet and \scalht bins. If the closure tests use
the \mmj sample, the level of closure is determined by additionally
integrating over pairs of adjacent \scalht bins. These uncertainties,
derived from the closure tests in data, are summarised in
Table~\ref{tab:bkgd_systs}, along with representative
magnitudes. These uncertainties are the dominant contribution to the
total uncertainty in the \tf factors, due to the limited number
of events in the data control regions.

As introduced in Section~\ref{sec:mht_templates}, templates are
derived from simulation to predict the \HTmiss distributions of the
background. The uncertainties in the \tf factors are used to constrain the
normalisation of the \HTmiss templates. The uncertainties in the
\HTmiss shape are discussed below.

The accuracy to which the simulation describes the \HTmiss
distributions is evaluated with respect to data in each (\njet, \nb,
\scalht) bin in each of the \mj, \mmj, and \gj data control regions.
The level of agreement between data and simulation, defined in terms
of the ratio of observed and expected counts (from simulation) as a
function of \HTmiss, is parameterised using an orthogonal first-order
polynomial, $f(x) = p_0 + p_1(\bar{x}-x)$, and described by two
uncorrelated parameters, $p_0$ and $p_1$. A binned likelihood fit is
performed in each (\njet, \nb, \scalht) bin of each control region,
and the best fit value $p_1$ and its uncertainty is used to determine
the presence of biases dependent on \HTmiss. The pull of $p_1$ from a
value of zero is defined as the best fit value over its standard
deviation, considering only statistical uncertainties associated with
the finite size of the data and simulated samples.

The lower bound of the final (open) bin in \HTmiss is not more than
800\GeV and is bounded from above by the upper bound of the \scalht
bin in question. The lower bound of the final \HTmiss bin is merged
with lower bins if fewer than ten events in the data control regions
are observed. If a bin in (\njet, \nb, \scalht) contains fewer than
ten events, the \HTmiss template is not used and the background
estimates are determined inclusively with respect to \HTmiss. The
merging of bins is typically only relevant for event categories that
satisfy $\nb \geq 2$.

The presence of systematic biases is evaluated at a statistical level
by considering the distribution of pulls obtained from each control
region, which are consistent with statistical fluctuations, with no
indication of trends across the full phase space of each control
region. The $p$-values obtained from the fits are uniformly
distributed.

The uncertainty in the \HTmiss modelling is extracted under the
hypothesis of no bias. This is done using the maximum likelihood (ML)
values of the fit parameters to determine the statistical precision to
which this hypothesis can be confirmed. The quadrature sum of the ML
value and its uncertainty for $p_1$ from each fit is used to define
alternative templates that represent $\pm1\sigma$ variations to the
nominal \HTmiss template. These alternative templates are encoded in
the likelihood function, as described in Section~\ref{sec:result}. The
observed variations are compatible with the expected values obtained
from studies relying only on simulated event samples. The
uncertainties in the final \HTmiss bin of the templates depend on the
event category and \scalht bin, and are typically found to be in the
range $\sim$10--100\%.

The effect on the \HTmiss templates is determined under $\pm1\sigma$
variations in the jet energy scale, the efficiency and
misidentification probability of b-quark jets, the efficiency to
identify well-reconstructed, isolated leptons, the pileup reweighting,
and the modelling of the top quark \pt. These effects
are easily covered by the uncertainties determined from data, as
described above, across the full phase space of the control regions,
which mirror closely that of the signal region.

\section{Results}
\label{sec:result}

A model of the observations in all data samples, described by a
likelihood function, is used to obtain a prediction of the SM
backgrounds and to test for the presence of new-physics signals if
the signal region is included in the ML fit. The
observation in each bin defined by the \njet, \nb, \scalht, and
\HTmiss variables is modelled as a Poisson-distributed variable around
the SM expectation and a potential signal contribution (assumed to be
zero in the following discussion), where the SM expectation is the sum
over the estimated contributions from all background processes
according to the methods described in Section~\ref{sec:backgrounds}.

The nonmultijet backgrounds
are related to the expected yields in the \mj, \mmj, and \gj control
samples via the transfer factors derived from simulation, as described
in Section~\ref{sec:control_regions}.
Estimates of the contribution from multijet events in the signal
region are determined according to the method described in
Section~\ref{sec:qcd_background}, and are included in the likelihood
function.

The systematic uncertainties summarised in Table~\ref{tab:bkgd_systs}
are accommodated in the likelihood function through the use of
nuisance parameters, the measurements of which are assumed to follow a
log-normal distribution. Alternative templates are used to describe
the uncertainties in the modelling of the \HTmiss variable. A vertical
template morphing scheme~\cite{Prosper:2011zz} is used to interpolate
between the nominal and alternative \HTmiss templates. A nuisance
parameter controls the interpolation, which is Gaussian distributed
with a mean of zero and a standard deviation of one, where $\pm$1
corresponds to the alternative templates for a $\pm$1$\sigma$
variation in the uncertainty. Each template is interpolated
quadratically between $\pm$1$\sigma$, and a linear extrapolation is
employed beyond these bounds.

\begin{figure*}[!thb]
  \centering
    \includegraphics[width=0.75\textwidth]{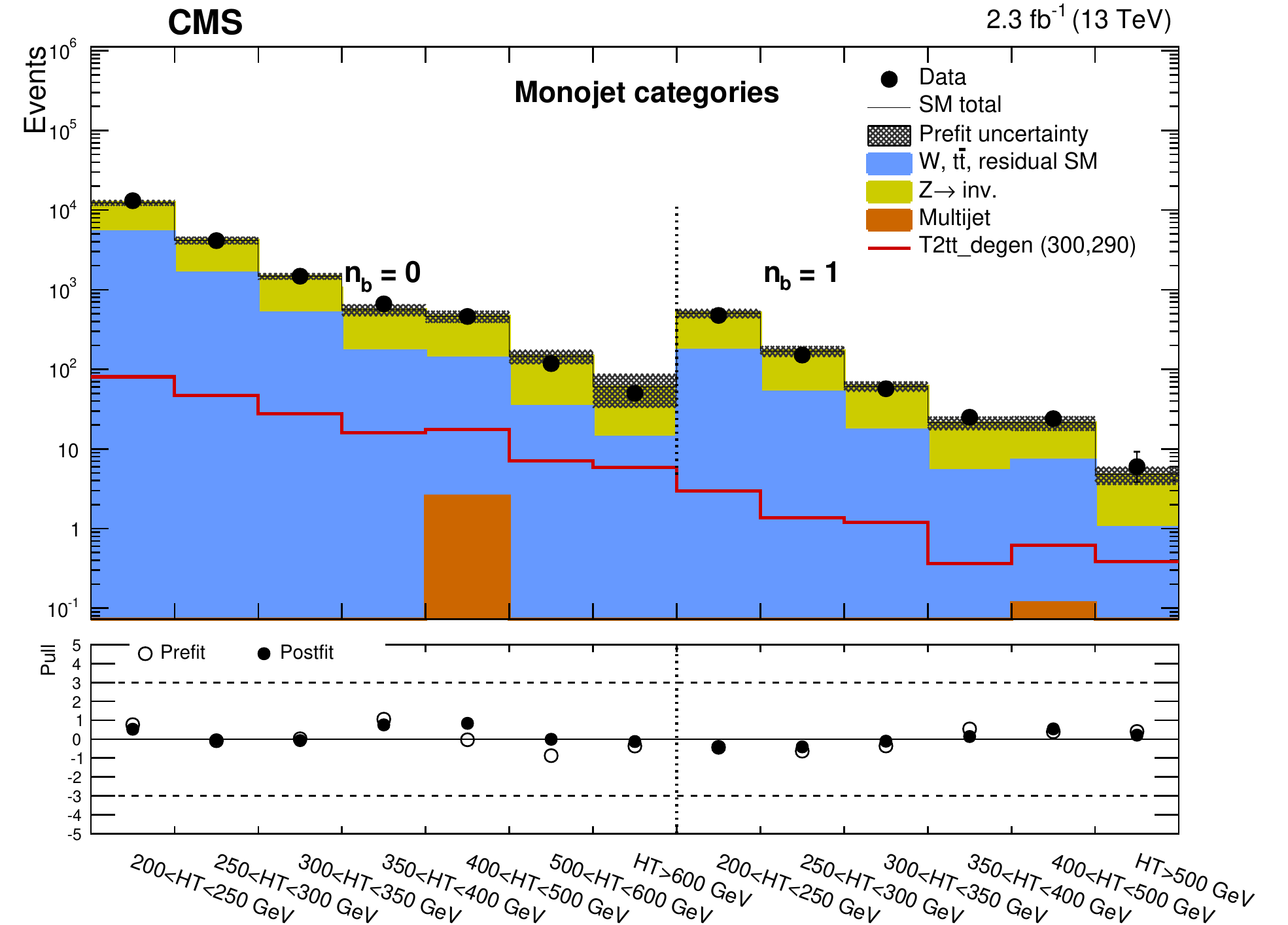}
    \caption{(Upper panel) Event yields observed in data (solid
      circles) and pre-fit SM expectations with their associated
      uncertainties (black histogram with shaded band), integrated
      over \HTmiss, as a function of \nb and \scalht for the monojet
      category ($\njet = 1$) in the signal region. For illustration
      only, the expectations for a benchmark model
      (\texttt{T2tt\_degen} with $m_{\,\PSQt} = 300\GeV$ and
      $m_{\PSGczDo} = 290\GeV$) are superimposed on the SM-only
      expectations. (Lower panel) The significance of deviations
      (pulls) observed in data with respect to the pre-fit (open
      circles) and post-fit (closed circles) SM expectations,
      expressed in terms of the total uncertainty in the SM
      expectations. The pulls cannot be considered independently due
      to inter-bin correlations.}
    \label{fig:mono}
\end{figure*}

\begin{figure*}[!hbt]
  \centering
    \includegraphics[width=\cmsFigWidthTwo]{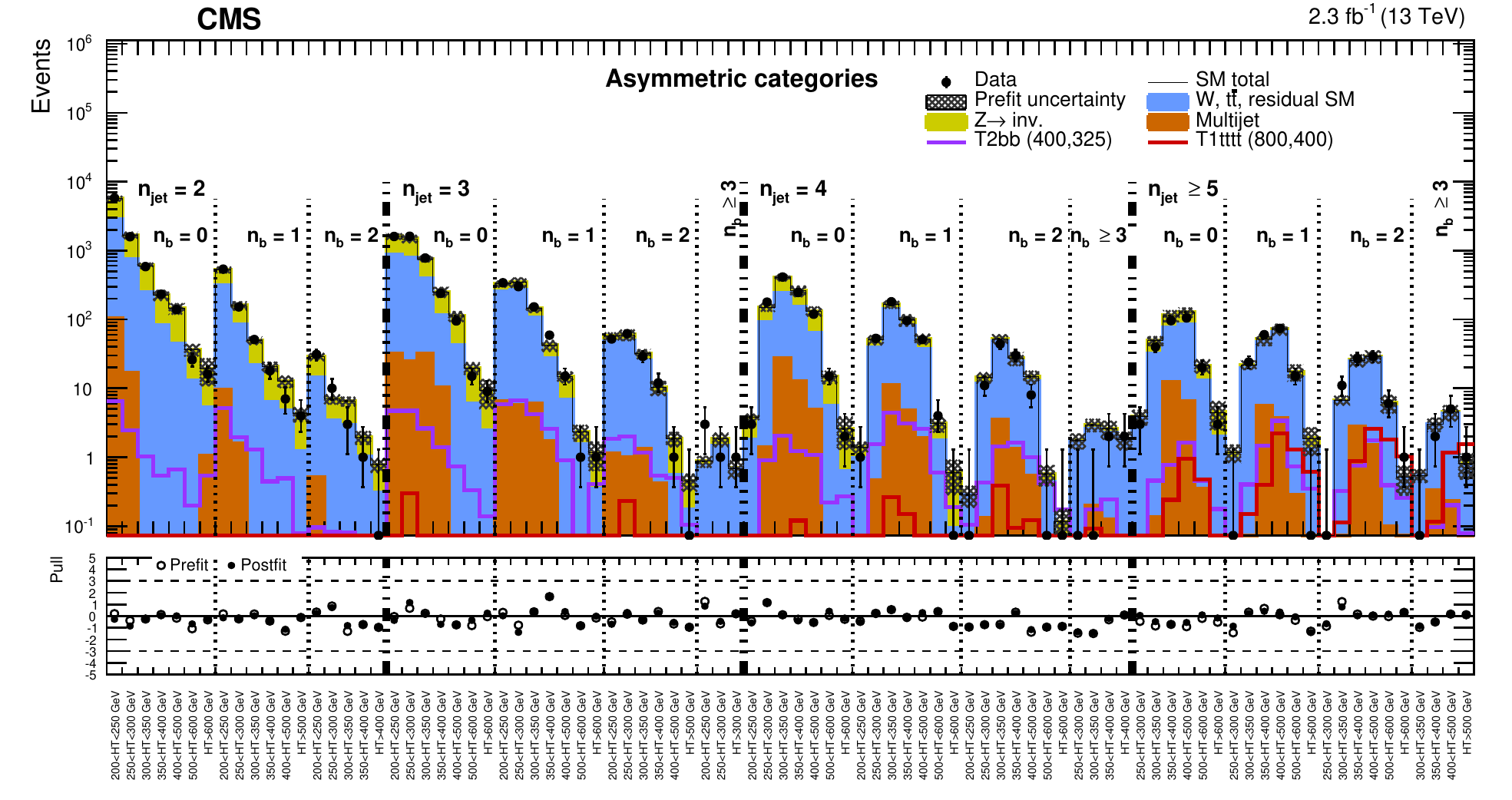}
    \caption{(Upper panel) Event yields observed in data (solid
      circles) and pre-fit SM expectations with their associated
      uncertainties (black histogram with shaded band), integrated
      over \HTmiss, as a function of \njet, \nb, and \scalht for the
      asymmetric \njet categories in the signal region. For
      illustration only, the expectations for two benchmark models
      (\texttt{T2bb} with $m_{\,\PSQb} = 400\GeV$ and $m_{\PSGczDo} =
      325\GeV$, \texttt{T1tttt} with $m_{\,\PSg} = 800\GeV$ and
      $m_{\PSGczDo} = 400\GeV$) are superimposed on the SM-only
      expectations. (Lower panel) The significance of deviations
      (pulls) observed in data with respect to the pre-fit (open
      circles) and post-fit (closed circles) SM expectations,
      expressed in terms of the total uncertainty in the SM
      expectations. The pulls cannot be considered independently due
      to inter-bin correlations.}
    \label{fig:asym}
\end{figure*}

\begin{figure*}[!htb]
  \centering
    \includegraphics[width=\cmsFigWidthTwo]{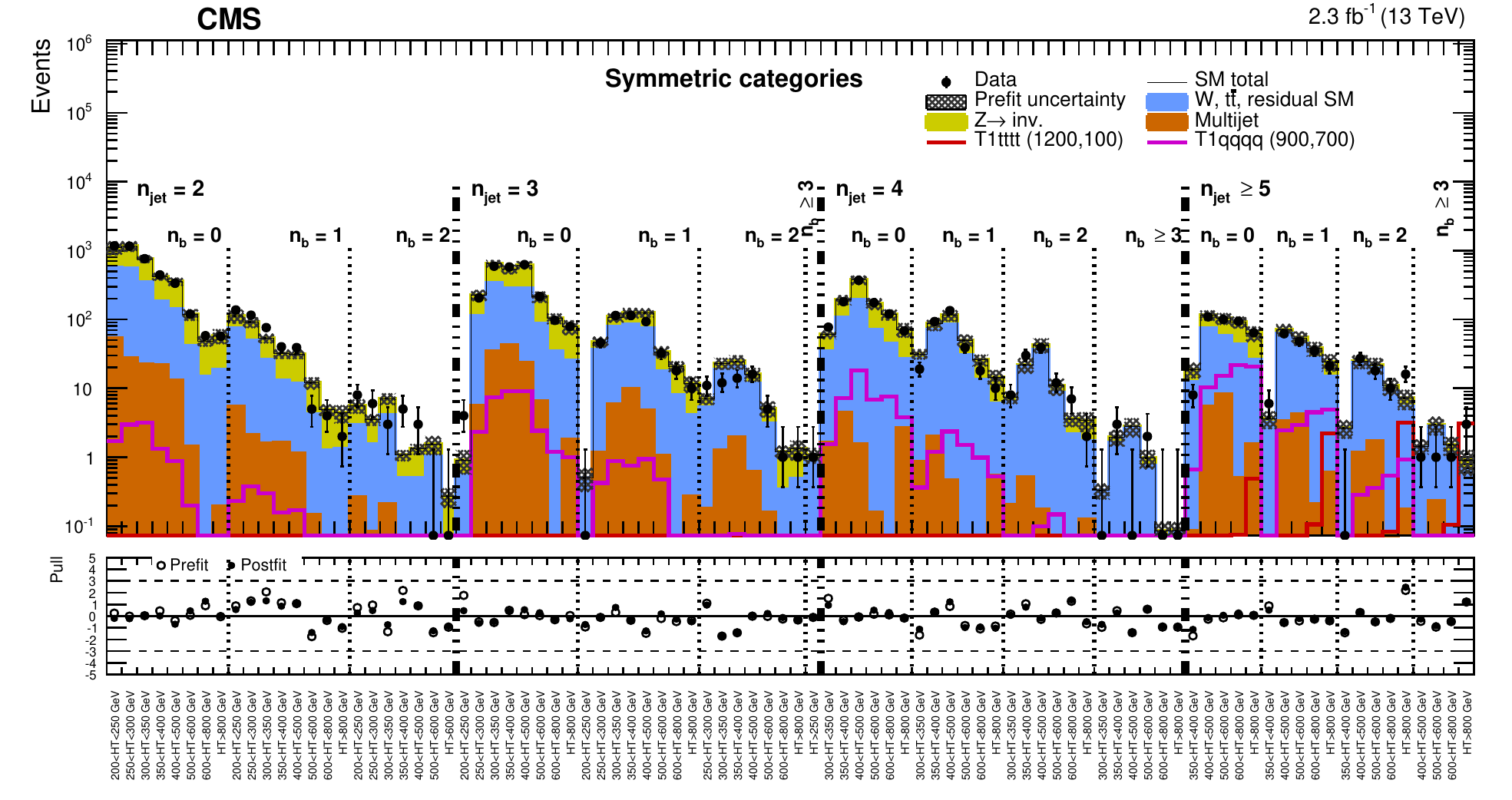}
    \caption{(Upper panel) Event yields observed in data (solid
      circles) and pre-fit SM expectations with their associated
      uncertainties (black histogram with shaded band), integrated
      over \HTmiss, as a function of \njet, \nb, and \scalht for the
      symmetric \njet categories in the signal region. For
      illustration only, the expectations for two benchmark models
      (\texttt{T1tttt} with $m_{\,\PSg} = 1200\GeV$ and $m_{\PSGczDo}
      = 100\GeV$, \texttt{T1qqqq} with $m_{\,\PSg} = 900\GeV$ and
      $m_{\PSGczDo} = 700\GeV$) are superimposed on the SM-only
      expectations. (Lower panel) The significance of deviations
      (pulls) observed in data with respect to the pre-fit (open
      circles) and post-fit (closed circles) SM expectations,
      expressed in terms of the total uncertainty in the SM
      expectations. The pulls cannot be considered independently due
      to inter-bin correlations.}
    \label{fig:sym}
\end{figure*}

The data are inspected to ascertain whether they are well described by
the null (SM-only) hypothesis. This is done by considering the
``pre-fit'' SM background estimates, which are determined from
observed data counts in the control regions only. The pre-fit result
of this search is summarised in Figs.~\ref{fig:mono}--\ref{fig:sym}
for, respectively, the monojet, asymmetric, and symmetric
topologies. The figures also show the significance of deviations
observed in data with respect to the pre-fit SM expectations expressed
in terms of the total uncertainty in the SM expectations
(``pull''). The data are well described by the background-only
hypothesis. Figures~\ref{fig:mono}--\ref{fig:sym} also summarise the
pulls from the post-fit result, which is based on a ML fit to
observations in the signal region as well as the control regions.

A quantitative statement on the degree of compatibility between the
observed yields and the SM expectations under the background-only
hypothesis is obtained from a goodness-of-fit test based on a log
likelihood ratio. The alternative hypothesis is defined by a
"saturated" model~\cite{sat-llk}, for which the background expectation
is set equal to the observed number of events, and provides a
reference for the largest value that the likelihood can take for any
model for the given data set. Hence, this reference can be used as a
reasonable normalisation for the maximum value observed for a more
constraining model. A $p$-value of 0.20 is observed for the fit over
the full signal region, and $p$-values in the range 0.04--1.00,
consistent with a uniform distribution, are obtained when considering
events categorised according to \njet.

The covariance and correlation matrices for the pre-fit SM
expectations in all bins of the signal region, defined by \njet, \nb,
\scalht, and integrated over \HTmiss, are determined from 500
pseudo-experiments by sampling the pre-fit nuisance parameters under
the background-only hypothesis. The SM expectations for different
\njet and \nb categories exhibit a nonnegligible level of covariance
within the same \scalht bin, primarily as a result of the systematic
uncertainties evaluated from closure tests, described in
Section~\ref{sec:ewk_background}, that integrate yields over \njet and
\nb. Bins adjacent and next-to-adjacent in \njet and/or \nb can have
correlation coefficients in the range 0.2--0.4, and, infrequently, as
large as $\sim$0.5. Otherwise, the correlation coefficients are
$<$0.2. Anticorrelation coefficients are typically not larger than
$\sim$0.2.

\begin{figure}[!tbh]
  \centering
    \includegraphics[width=0.49\textwidth]{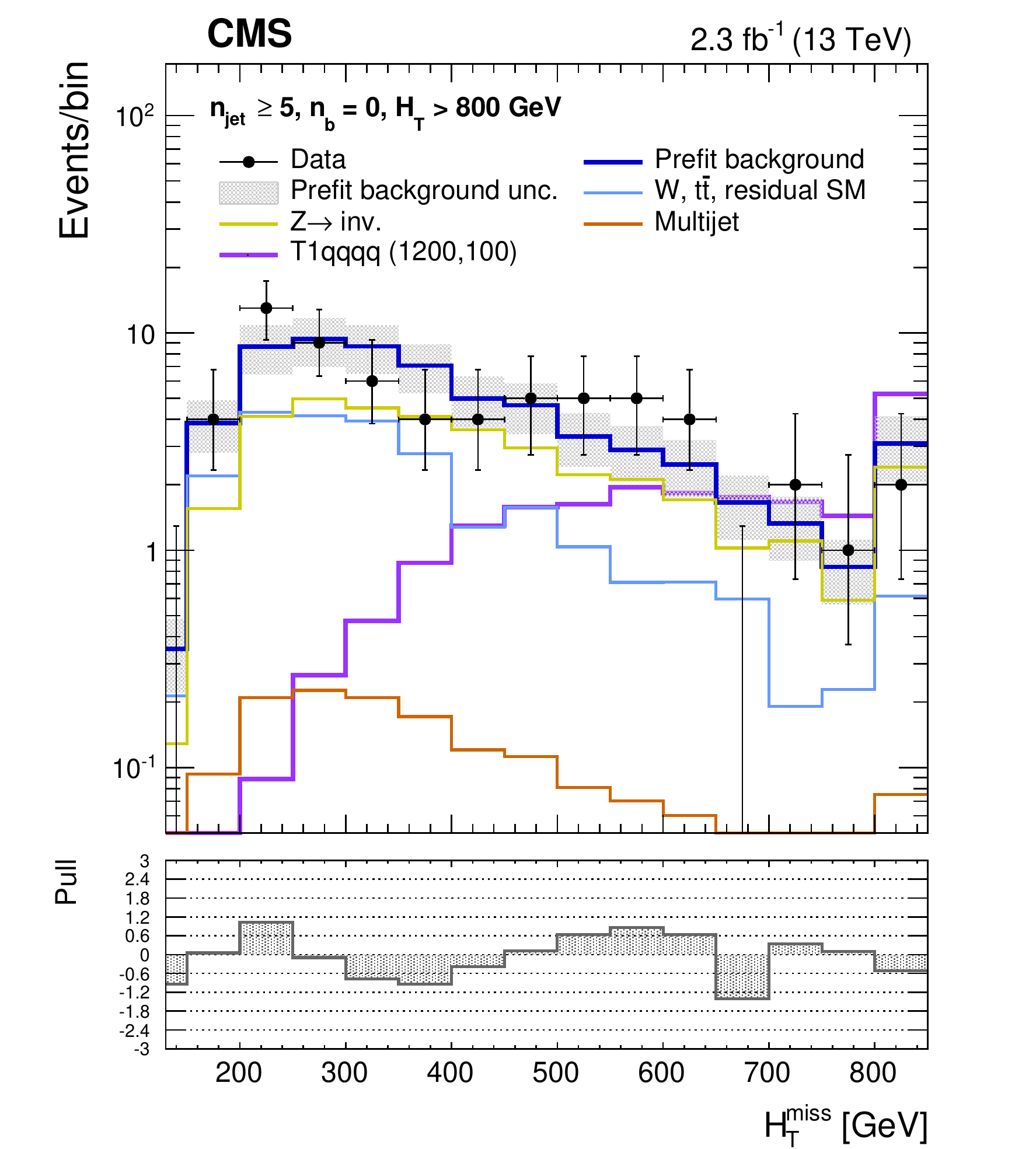}
    \includegraphics[width=0.49\textwidth]{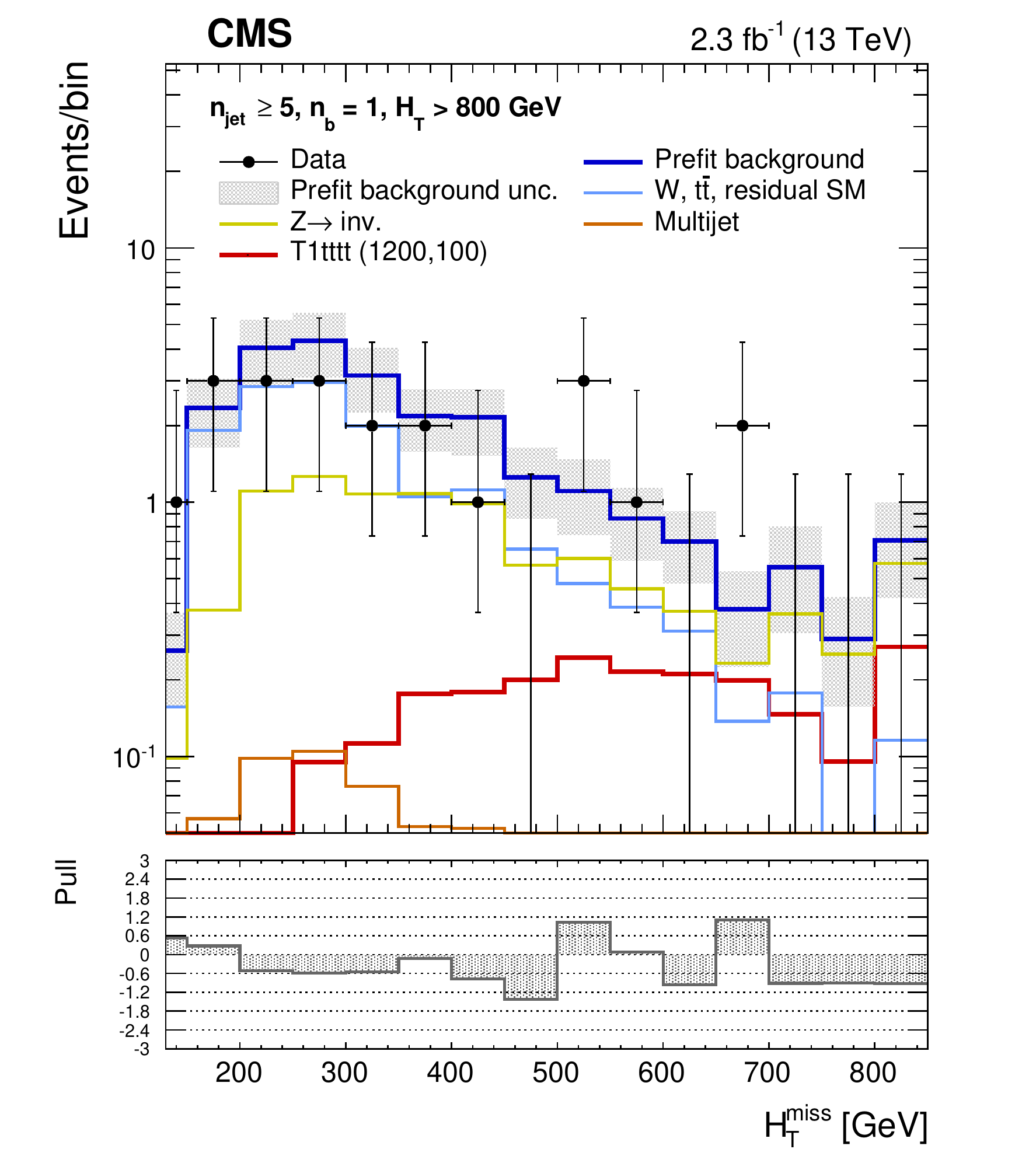} \\
    \caption{Event yields observed in data (solid circles) and pre-fit
      SM expectations with their associated uncertainties (blue
      histogram with shaded band) as a function of \HTmiss for events
      in the signal region that satisfy $\njet \geq 5$, $\scalht >
      800\GeV$, and (\cmsLeft) $\nb = 0$ or (\cmsRight) $\nb = 1$. For
      illustration only, the expectations for one of two benchmark
      models (\texttt{T1qqqq} and \texttt{T1tttt}, both with
      $m_{\,\PSg} = 1200\GeV$ and $m_{\PSGczDo} = 100\GeV$) are
      superimposed on the SM-only expectations. The lower panels
      indicate the significance of deviations (pulls) observed in data
      with respect to both the pre-fit SM expectations, expressed in
      terms of the total uncertainty in the SM expectations. The pulls
      cannot be considered independently due to inter-bin
      correlations. }
    \label{fig:mht-templates}
\end{figure}

Figure~\ref{fig:mht-templates} shows the event yields observed in data
and pre-fit SM expectations with their associated uncertainties as
a function of \HTmiss for two categories of candidate signal events,
which provide good sensitivity to models with high-mass gluinos. For
illustration only, the expected counts from benchmark signal models
that assume the pair production and decay of gluinos, described
further in Section~\ref{sec:interpretations}, are also shown.

\section{Interpretation}
\label{sec:interpretations}

\subsection{Specification for simplified models}

The results of the search are used to constrain simplified SUSY
models~\cite{Alwall:2008ag, Alwall:2008va, sms}. Each model assumes
the pair production of gluinos or squarks and their subsequent prompt
decays to SM particles and the LSP with a 100\% branching fraction
(unless indicated otherwise). The gluino decays contain intermediate
on-shell SUSY particle states (such as the top squark or the chargino)
for a subset of the models. All other SUSY particles are assumed to be
too heavy ($m_{\PSg} / m_{\PSQ} = 10\TeV$) to be produced
directly. Three-body decays of gluinos are assumed to occur via
off-shell squarks of light or heavy flavour. Off-shell decays are
processed by \PYTHIA in a single three- or four-body step, without
taking into account the width or polarisation of the parent: this is
true for the top-squark four-body decay ($\PSQt \to
\mathrm{bf}\overline{\mathrm{f}}'\PSGczDo$), as well as the three-body decay of
the chargino ($\PSGcpm_1 \to  \text{f}\bar{\text{f}}'\PSGczDo$), where f
and f$'$ are fermions produced in the decay of an intermediate
off-shell W boson.

\begin{table*}[!bt]
  \topcaption{A summary of the simplified SUSY models used
    to interpret the results of this search. All on-shell SUSY particles
    in the decay are stated.}
  \label{tab:simplified-models}
  \centering
  \begin{tabular}{ llll }
    \hline
Model class
                                & Production
                                & Decay
                                & Additional assumptions                                                         \\ [0.5ex]
\hline
\multicolumn{4}{l}{Gluino-mediated and direct production of light-flavour squarks}                           \\ [1ex]
\texttt{T1qqqq}
                                & $\Pp\Pp\to \PSg\PSg$
                                & $\PSg\to \cPaq\cPq\PSGczDo$
                                & \NA                                                                             \\ [0.5ex]
\texttt{T2qq\_8fold}
                                & $\Pp\Pp\to \PSQ\PASQ$
                                & $\PSQ\to \cPq\PSGczDo$
                                & $m_{\PSQ} = m_{\PSQ_\cmsSymbolFace{L}} = m_{\PSQ_\cmsSymbolFace{R}}$,
                       $\PSQ = \{ \PSQu, \PSQd, \PSQs, \PSQc \}$                                                 \\ [0.5ex]
\texttt{T2qq\_1fold}
                                & $\Pp\Pp\to \PSQ\PASQ$
                                & $\PSQ\to \cPq\PSGczDo$
                                & $m_{\PSQ (\PSQ \neq \PSQu_\cmsSymbolFace{L})} \gg m_{\PSQu_\cmsSymbolFace{L}}$ \\ [2.5ex]
\multicolumn{4}{l}{Gluino-mediated production of off-shell third-generation squarks}                         \\ [1ex]
\texttt{T1bbbb}
                                & $\Pp\Pp\to \PSg\PSg$
                                & $\PSg\to \cPaqb\cPqb\PSGczDo$
                                & \NA                                                                             \\ [0.5ex]
\texttt{T1tttt}
                                & $\Pp\Pp\to \PSg\PSg$
                                & $\PSg\to \cPaqt\PSQt^*\to \cPaqt\cPqt\PSGczDo$
                                & \NA                                                                             \\ [0.5ex]
\texttt{T1ttbb}
                                & $\Pp\Pp\to \PSg\PSg$
                                & $\PSg\to \cPaqt\cPqb\PSGcpm_1\to \cPaqt\cPqb\PW^*\PSGczDo$
                                & $m_{\PSGcpm_1} - m_{\PSGczDo} = 5\GeV$                                           \\ [2.5ex]
\multicolumn{4}{l}{Natural gluino-mediated production of on-shell top squarks}                               \\ [1ex]
\texttt{T5tttt\_DM175}
                                & $\Pp\Pp\to \PSg\PSg$
                                & $\PSg\to \cPaqt\PSQt\to \cPaqt\cPqt\PSGczDo$
                                & $m_{\,\PSQt} - m_{\PSGczDo} = 175\GeV$                                          \\ [0.5ex]
\texttt{T5ttcc}
                                & $\Pp\Pp\to \PSg\PSg$
                                & $\PSg\to \cPaqt\PSQt\to \cPaqt\cPqc\PSGczDo$
                                & $m_{\,\PSQt} - m_{\PSGczDo} = 20\GeV$                                           \\ [2.5ex]
\multicolumn{4}{l}{Direct production of on-shell third-generation squarks}                                   \\ [1ex]
\texttt{T2bb}
                                & $\Pp\Pp\to \PSQb\PASQb$
                                & $\PSQb\to \cPqb\PSGczDo$
                                & \NA                                                                             \\ [0.5ex]
\texttt{T2tb}
                                & $\Pp\Pp\to \PSQt\PASQt$
                                & $\PSQt\to \cPqt\PSGczDo \;\text{or}\; \cPqb\PSGcpm_1\to \cPqb\PW^*\PSGczDo$
                                & $50/50\%$, $m_{\PSGcpm_1} - m_{\PSGczDo} = 5\GeV$                                \\ [0.5ex]
\texttt{T2tt}
                                & $\Pp\Pp\to \PSQt\PASQt$
                                & $\PSQt\to \cPqt\PSGczDo$
                                & \NA                                                                             \\ [0.5ex]
\texttt{T2cc}
                                & $\Pp\Pp\to \PSQt\PASQt$
                                & $\PSQt\to \cPqc\PSGczDo$
                                & $10 < m_{\,\PSQt} - m_{\PSGczDo} < 80\GeV$                                      \\ [0.5ex]
\texttt{T2tt\_degen}
                                & $\Pp\Pp\to \PSQt\PASQt$
                                & $\PSQt\to \cPqb\PW^*\PSGczDo$
                                & $10 < m_{\,\PSQt} - m_{\PSGczDo} < 80\GeV$                                      \\ [0.5ex]
\texttt{T2tt\_mixed}
                                & $\Pp\Pp\to \PSQt\PASQt$
                                & $\PSQt\to \cPqc\PSGczDo \;\text{or}\; \cPqb\PW^*\PSGczDo$
                                & $50/50\%$, $10 < m_{\,\PSQt} - m_{\PSGczDo} < 80\GeV$                           \\ [0.5ex]
    \hline
  \end{tabular}
\end{table*}

Fourteen unique production and decay modes are considered, which yield
a range of topologies and final states (with only the all-jet final
state considered in this search). Each class of simplified model is
identified by a label that indicates the topology and final state, and
scans in the gluino or squark ($m_{\PSg} / m_{\PSQ}$) and LSP
($m_{\PSGczDo}$) mass parameter space are performed.
Table~\ref{tab:simplified-models} summarises the production and decay
modes, as well as any additional assumptions that define the
simplified models. The models can be categorised according to the
following descriptions: the gluino-mediated and direct production of
light-flavour squarks, the gluino-mediated production of off-shell
third-generation squarks, the natural gluino-mediated production
of on-shell top squarks, and the direct production of on-shell
third-generation squarks. In the case of direct pair production of
light-flavour squarks, two different assumptions on the theoretical
production cross section are made. For the ``eightfold'' scenario
(\texttt{T2qq\_8fold}), the scalar partners to left- and right-handed
quarks of the u, d, s, and c flavours are assumed to be light and
degenerate in mass, with other squark states decoupled to a high
mass. For the ``onefold'' scenario (\texttt{T2qq\_1fold}), only a
single light squark is assumed to participate in the interaction and
all other squarks are decoupled to a high mass.

Under the signal+background hypothesis, and in the presence of a
nonzero signal contribution, a modified frequentist approach is used
to determine upper limits at the 95\% confidence level (CL) on the
cross section, $\sigma_\mathrm{UL}$, to produce pairs of SUSY
particles as a function of the parent SUSY particle and the LSP
masses. The limits can be expressed in terms of the signal strength
parameter, $\mu$, which is determined relative to the theoretical
cross section that is calculated at NLO+NLL accuracy. 
An Asimov data set~\cite{Cowan:2010js} is used to determine the
expected upper limit on the allowed cross section for a given model.
The potential contributions from a new-physics signal to each of the
signal and control regions are considered, even though the only
significant contribution occurs in the signal region and not in the
control regions (\ie signal contamination). The approach is based on
the one-sided (so called LHC-style) profile likelihood ratio as the
test statistic~\cite{CMS-NOTE-2011-005} and the \cls
criterion~\cite{junk, read}. Asymptotic formulae~\cite{Cowan:2010js}
are utilised to approximate the distributions of the test statistics
under the SM background-only and signal+background hypotheses. 

\begin{table*}[!t]
  \topcaption{A summary of benchmark simplified models, the most sensitive
    \njet categories, and representative values for the corresponding
    experimental acceptance times efficiency
    ($\mathcal{A}\,\varepsilon$), the dominant systematic
    uncertainties, the theoretical production cross section
    ($\sigma_\text{theory}$), and the expected and observed upper limits
    on the production cross section, expressed in terms of the signal
    strength parameter ($\mu$).
  }
  \label{tab:signal-eff}
  \centering
  \resizebox{\textwidth}{!}{
    \begin{tabular}{ lllcrrrrrcc }
      \hline
      \multicolumn{2}{l}{Benchmark models}
 & Most sensitive
 & $\mathcal{A}\,\varepsilon$
 & \multicolumn{4}{c}{Systematic uncertainties [\%]}
 & $\sigma_\text{theory}$
 & \multicolumn{2}{c}{$\mu$ (95\% CL)}                                                                             \\ [0.3ex]
      \cline{5-8}
      \multicolumn{2}{l}{$(m_{\text{SUSY}}, m_{\mathrm{LSP}})$ [\GeVns{}]}
 & \njet categories
 & [\%]
 & MC stat.
 & ISR
 & JEC
 & $\text{SF}_\text{b-tag}$
 & \multicolumn{1}{c}{[fb]}
 & Exp.
 & Obs.                                                                                                            \\ [0.3ex]
      \hline
      \multirow{2}{*}{\texttt{T1qqqq}}
 & (1300, 100) & $\geq$5, 4, 3, 2         & 21.2           & 7--30  & $\sim$2 & 4--21 & 2--14 & 46.1 & 0.79 & 0.76 \\
 & (900, 700)  & $\geq$5, $\geq$5a, 4, 4a & 12.8           & 10--33 & 1--13   & 1--26 & 1--10 & 677  & 0.58 & 0.44 \\ [0.5ex]
    \multirow{2}{*}{\texttt{T2qq\_8fold}}
 & (1050, 100) & $\geq$5, 3, 4, 2         & 40.3           & 7--33  & 1--4    & 3--16 & 1--11 & 35.2 & 0.90 & 0.63 \\
 & (650, 550)  & $\geq$5, 4, $\geq$5a, 4a & \phantom{1}6.3 & 10--28 & 1--16   & 2--29 & 1--6  & 864  & 0.93 & 0.80 \\ [0.5ex]
    \multirow{2}{*}{\texttt{T2qq\_1fold}}
 & (600, 50)   & $\geq$5, 3, 2, 4         & 30.2           & 5--33  & 1--5    & 1--30 & 1--8  & 177  & 0.78 & 0.84 \\
 & (400, 250)  & $\geq$5, 4, $\geq$5a, 3  & \phantom{1}7.1 & 8--30  & 1--8    & 3--25 & 1--7  & 1849 & 0.73 & 0.71 \\ [0.5ex]
      \multirow{2}{*}{\texttt{T1bbbb}}
 & (1500, 100) & $\geq$5, 4, 3, 2         & 22.7           & 5--17  & 1--2    & 1--12 & 2--22 & 14.2 & 0.81 & 0.79 \\
 & (1000, 800) & $\geq$5, 4, $\geq$5a, 4a & 11.4           & 8--31  & 1--17   & 1--40 & 1--14 & 325  & 0.33 & 0.32 \\ [0.5ex]
      \multirow{2}{*}{\texttt{T1tttt}}
 & (1300, 100) & $\geq$5, $\geq$5a, 4, 3  & \phantom{1}5.3 & 7--16  & 1--2    & 2--7  & 2--12 & 46.1 & 1.00 & 1.89 \\
 & (800, 400)  & $\geq$5, $\geq$5a, 4, 4a & \phantom{1}1.5 & 7--27  & 1--2    & 3--45 & 1--8  & 1490 & 0.56 & 1.03 \\ [0.5ex]
      \multirow{2}{*}{\texttt{T1ttbb}}
 & (1300, 100) & $\geq$5, 4, 3, $\geq$5a  & \phantom{1}8.5 & 9--32  & 1--2    & 3--16 & 2--19 & 46.1 & 0.60 & 0.91 \\
 & (1000, 700) & $\geq$5, $\geq$5a, 4, 3  & \phantom{1}7.7 & 9--30  & 1--9    & 3--65 & 1--14 & 325  & 0.51 & 0.70 \\ [0.5ex]
      \multirow{2}{*}{\texttt{T5tttt\_DM175}}
 & (800, 100)  & $\geq$5, $\geq$5a, 3, 4  & \phantom{1}0.5 & 12--20 & 2--4    & 3--5  & 1--6  & 1490 & 0.69 & 1.19 \\
 & (700, 400)  & $\geq$5, $\geq$5a, 4, 4a & \phantom{1}0.5 & 20--29 & 2--10   & 8--10 & 1--2  & 3530 & 1.00 & 1.35 \\ [0.5ex]
      \multirow{2}{*}{\texttt{T5ttcc}}
 & (1200, 200) & $\geq$5, 4, 3, $\geq$5a  & 11.0           & 6--25  & 5--25   & 3--21 & 1--24 & 85.6 & 0.58 & 0.87 \\
 & (750, 600)  & $\geq$5, $\geq$5a, 4, 4a & \phantom{1}2.2 & 9--23  & 1--4    & 5--21 & 1--3  & 2270 & 0.89 & 0.72 \\ [0.5ex]
      \multirow{2}{*}{\texttt{T2bb}}
 & (800, 50)   & 2, 3, 4, $\geq$5         & 34.9           & 5--31  & 2--6    & 1--21 & 1--23 & 28.3 & 0.96 & 1.06 \\
 & (375, 300)  & $\geq$5, 4, 3a, 3        & \phantom{1}3.2 & 8--33  & 1--10   & 3--25 & 1--7  & 2610 & 0.67 & 0.87 \\ [0.5ex]
      \multirow{2}{*}{\texttt{T2tb}}
 & (600, 50)   & $\geq$5, 4, 3, 2         & 13.4           & 3--28  & 1--3    & 1--22 & 1--17 & 175  & 0.70 & 1.35 \\
 & (350, 225)  & $\geq$5, 4, 3, 3a        & \phantom{1}2.3 & 9--33  & 1--4    & 2--41 & 1--8  & 3790 & 0.79 & 0.88 \\ [0.5ex]
      \multirow{2}{*}{\texttt{T2tt}}
 & (700, 50)   & $\geq$5, 4, 3, $\geq$5a  & 18.2           & 8--33  & 1--4    & 2--22 & 1--21 & 67.0 & 0.90 & 1.19 \\
 & (350, 100)  & $\geq$5, $\geq$5a, 4a, 4 & \phantom{1}3.4 & 7--31  & 1--1    & 1--28 & 1--7  & 3790 & 0.44 & 0.50 \\ [0.5ex]
      \multirow{1}{*}{\texttt{T2cc}}
 & (325, 305)  & $\geq$5, 4, 3, 2         & \phantom{1}1.9 & 3--32  & 1--27   & 1--27 & 1--12 & 5600 & 0.92 & 0.68 \\ [0.5ex]
      \multirow{1}{*}{\texttt{T2tt\_degen}}
 & (300, 290)  & 3, 4, $\geq$5, 2         & \phantom{1}2.0 & 2--27  & 1--27   & 1--25 & 1--12 & 8520 & 0.56 & 0.41 \\ [0.5ex]
      \multirow{1}{*}{\texttt{T2tt\_mixed}}
 & (300, 250)  & $\geq$5, 4, $\geq$5a, 4a & \phantom{1}1.0 & 3--33  & 1--27   & 1--33 & 1--13 & 8520 & 0.99 & 0.58 \\ [0.5ex]
      \hline
    \end{tabular}
  }
\end{table*}

\subsection{Acceptances and uncertainties}

The experimental acceptance times efficiency
($\mathcal{A}\,\varepsilon$) and its uncertainty are evaluated
independently for each model class as a function of ($m_\text{SUSY},
m_\text{LSP}$). Table~\ref{tab:signal-eff} summarises
$\mathcal{A}\,\varepsilon$ for a number of benchmark models for
which the search yields an expected exclusion ($\mu \lesssim 1$). For
each topology, typically two different pairs of parent SUSY particle and
LSP masses ($m_\text{SUSY}, m_\text{LSP}$) are chosen that are
characterised by a large and a small (\ie compressed) difference
in parent SUSY particle and LSP masses. The four most sensitive event
categories, defined in terms of \njet, are used to determine
$\sigma_\text{UL}$. The categories used per benchmark model are listed
in Table~\ref{tab:signal-eff}, along with
$\mathcal{A}\,\varepsilon$ determined for these four categories.

The effects of several sources of uncertainty on
$\mathcal{A}\,\varepsilon$, as well as the potential for event
migration between bins of the signal region, are considered. The
potential effect of each source of uncertainty is assessed by
including in the likelihood function the alternative normalisations
and shapes for the \HTmiss templates with respect to the nominal
versions. The nominal and alternative templates are obtained from
simulated event samples for the signal models, and the alternative
templates effectively propagate the various input uncertainties to
determine their effects on the \njet, \nb, \scalht, and \HTmiss
distributions for the signal model.

In addition to the uncertainty in the integrated luminosity of
2.7\%~\cite{CMS:2016eto}, the following sources of uncertainty are
dominant: the statistical uncertainty arising from the finite size of
simulated signal samples, the modelling of ISR, the jet energy
corrections (JEC) evaluated in simulation, and the modelling of scale
factors applied to simulated event samples that correct for
differences in the efficiency and misidentification probability of
b-quark jets ($\text{SF}_\text{b-tag}$). The magnitude of each
contribution depends on the model and the masses of the parent
SUSY particle and LSP.

The $\mathcal{A}\,\varepsilon$ for models with small mass
splittings (\eg $m_{\PSQ} - m_{\PSGczDo} \lesssim m_{\text t}$) is due
largely to ISR, the modelling of which is evaluated by comparing the
simulated and measured \pt spectra of the system recoiling against the
ISR jets in \ttbar events, using the technique described in
Ref.~\cite{single-lepton-stop}. The uncertainty can be as large as
$\sim$30\%, and is the dominant systematic uncertainty for systems
with a compressed mass spectrum. Uncertainties in the jet energy
scale, as large as $\sim$40\%, can also be dominant for models
characterised by high jet multiplicities in the final state. The
uncertainties in $\text{SF}_\text{b-tag}$ can be as large as
$\sim$25\%. Table~\ref{tab:signal-eff} summarises these dominant
contributions to the uncertainty in $\mathcal{A}\,\varepsilon$ for
a range of benchmark models. Characteristic values for each model are
expressed in terms of a range that is representative of the values
across all bins of the signal region. The upper bound for each range
may be subject to moderate statistical fluctuations.

Further uncertainties with subdominant contributions are considered on
a similar footing. The uncertainties in the efficiency of identifying
well-reconstructed, isolated leptons are considered, with a typical
magnitude of $\sim$5\% and treated as anticorrelated between the
signal and control regions. The uncertainty of 5\% in
$\sigma_\text{in}$ is propagated through to the reweighting procedure
to account for differences between the simulated and measured
pileup. Finally, uncertainties in the simulation modelling of the
efficiencies of the trigger strategy employed by the search are
typically $<$10\%.

The choice of PDF set, or variations therein, predominantly affects
$\mathcal{A}\,\varepsilon$ through changes in the \pt spectrum of
the system recoil, which is covered by the ISR uncertainty, hence no
additional uncertainty is adopted. Uncertainties in
$\mathcal{A}\,\varepsilon$ due to variations in the
renormalisation and factorisation scales are determined to be
$\sim$5\%. In both cases, contributions to the uncertainty in the
theoretical production cross section are considered.

\subsection{Cross section and mass exclusions}

Limits for each of the aforementioned benchmark models are summarised
in Table~\ref{tab:signal-eff}, expressed in terms of $\mu$. All
benchmark models are expected to be excluded. The observed limits
fluctuate around the expected $\mu$ values, with some models
exhibiting a moderately weaker than expected limit.

\begin{figure}[htb!]
  \centering
    \includegraphics[width=0.49\textwidth]{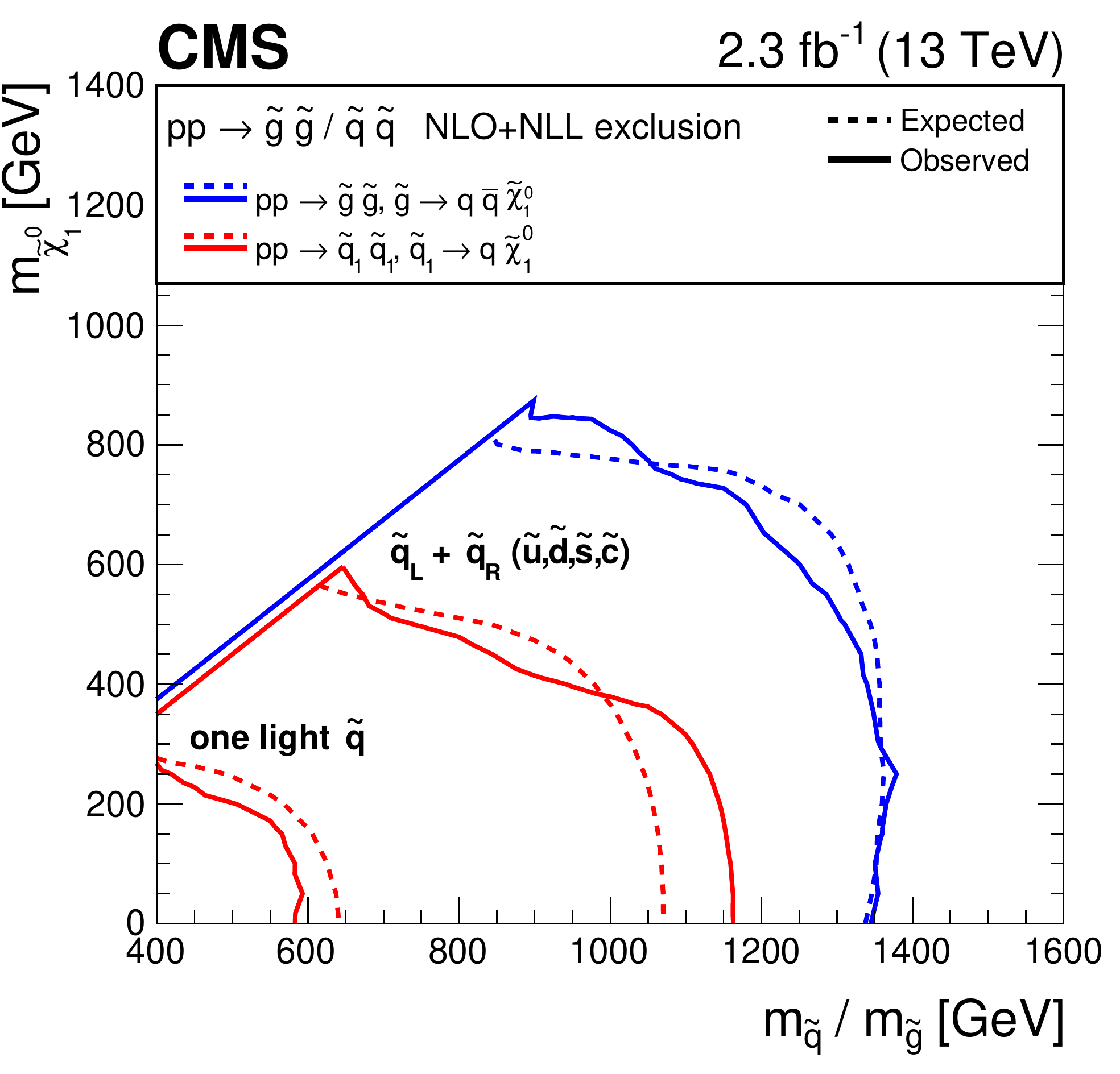}
    \includegraphics[width=0.49\textwidth]{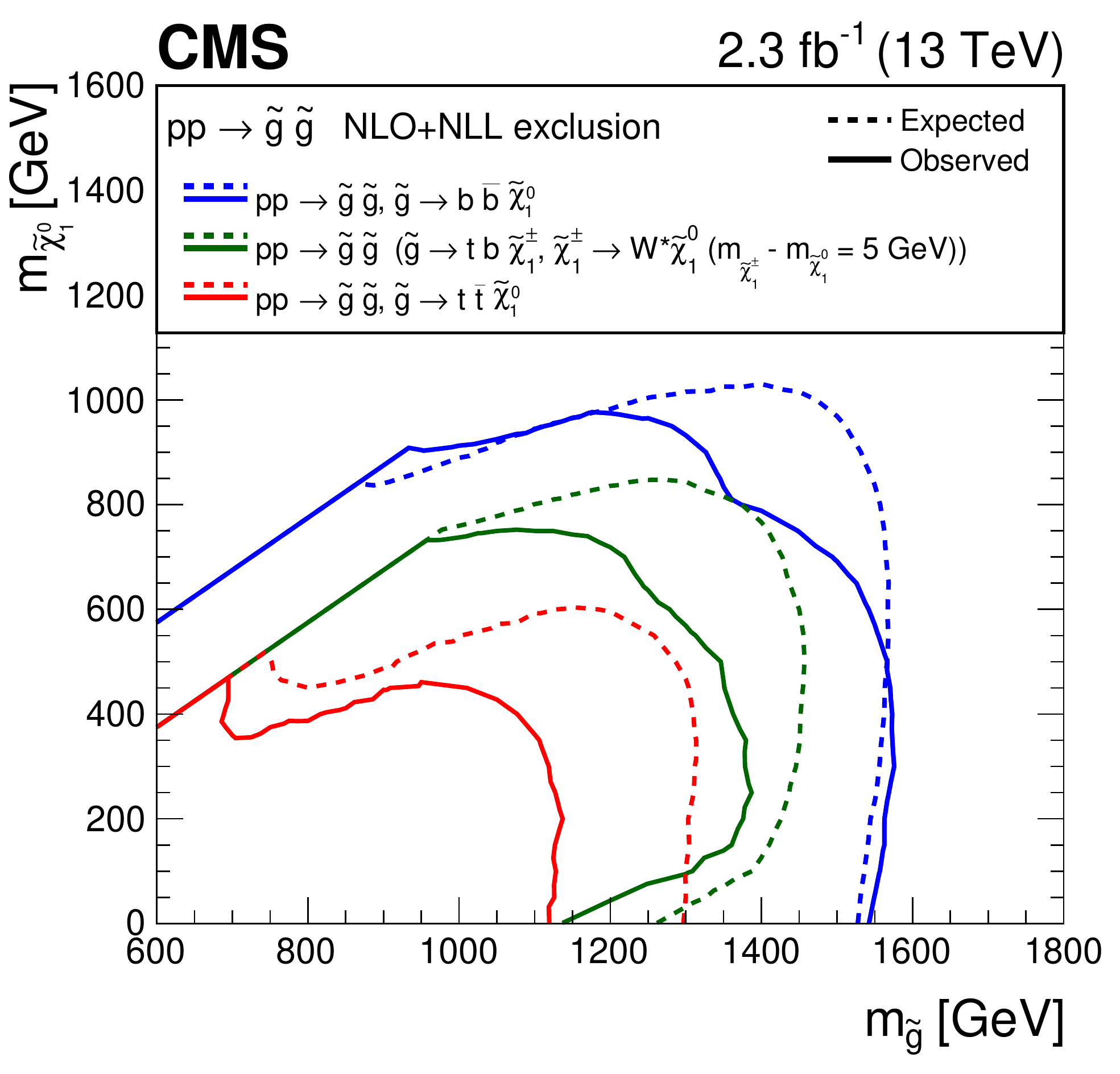}
    \caption{Observed and expected mass exclusions at 95\% CL
      (indicated, respectively, by solid and dashed contours) for
      various classes of simplified models. (\cmsLeft) Gluino-mediated or
      direct pair production of light-flavour squarks. The two
      scenarios involve, respectively, the decay
      $\PSg\to \cPaq\cPq\PSGczDo$ (\texttt{T1qqqq}) and
      $\PSQ\to \cPq\PSGczDo$, and the latter involves two assumptions on
      the mass degeneracy of the squarks (\texttt{T2qq\_8fold} and
      \texttt{T2qq\_1fold}). (\cmsRight) Three scenarios involving the
      gluino-mediated pair production of off-shell third-generation
      squarks: $\PSg\to \cPaqb\cPqb\PSGczDo$ (\texttt{T1bbbb}),
      $\PSg\to \cPaqt\PSQt^*\to \cPaqt\cPqt\PSGczDo$ (\texttt{T1tttt}),
      and $\PSg\to \cPaqt\cPqb\PSGcpm_1\to \cPaqt\cPqb\PW^*\PSGczDo$
      (\texttt{T1ttbb}).  }
    \label{fig:limits-sms-1}
\end{figure}

\begin{figure}[htb!]
  \centering
    \includegraphics[width=0.49\textwidth]{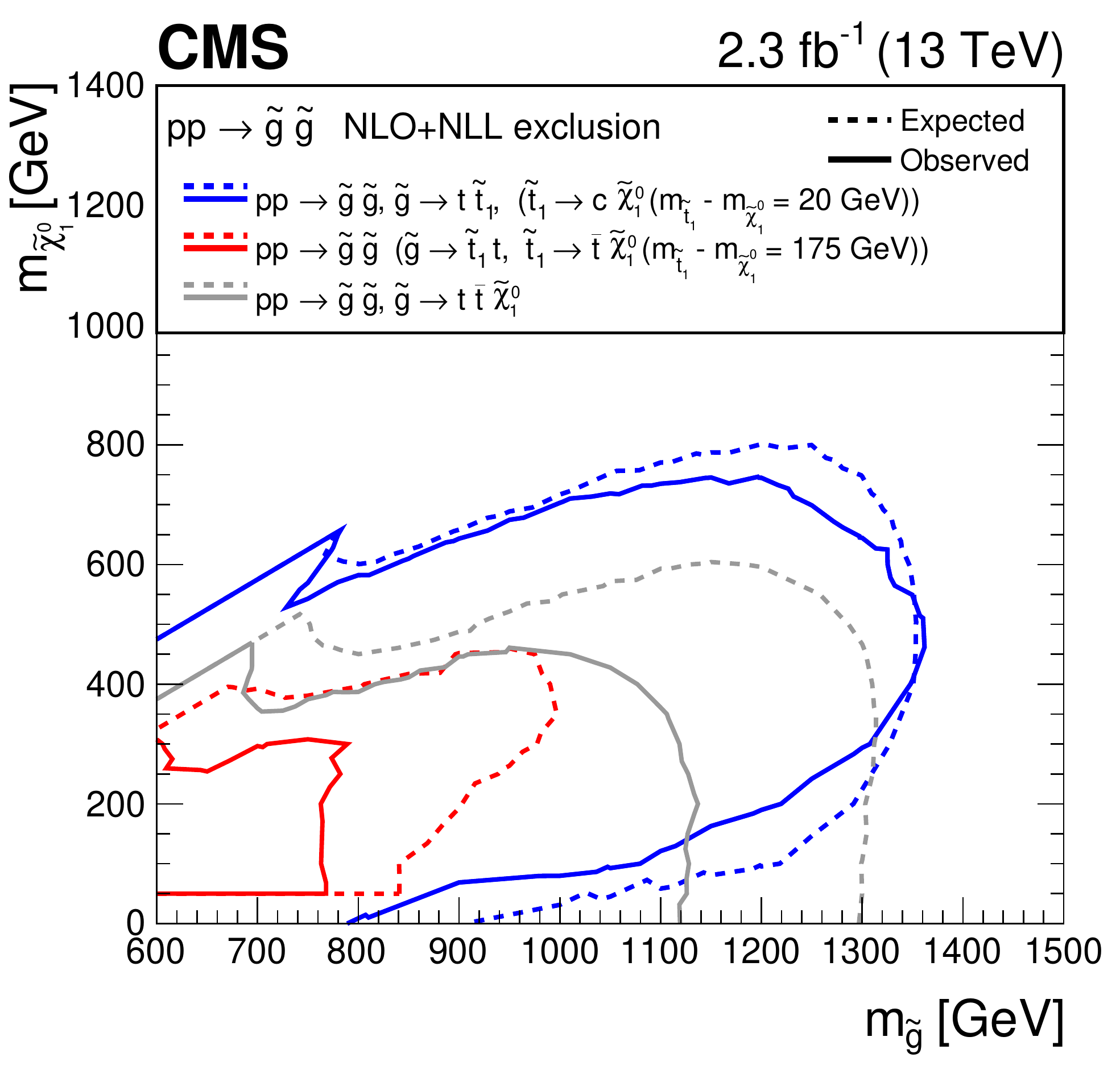}
    \includegraphics[width=0.49\textwidth]{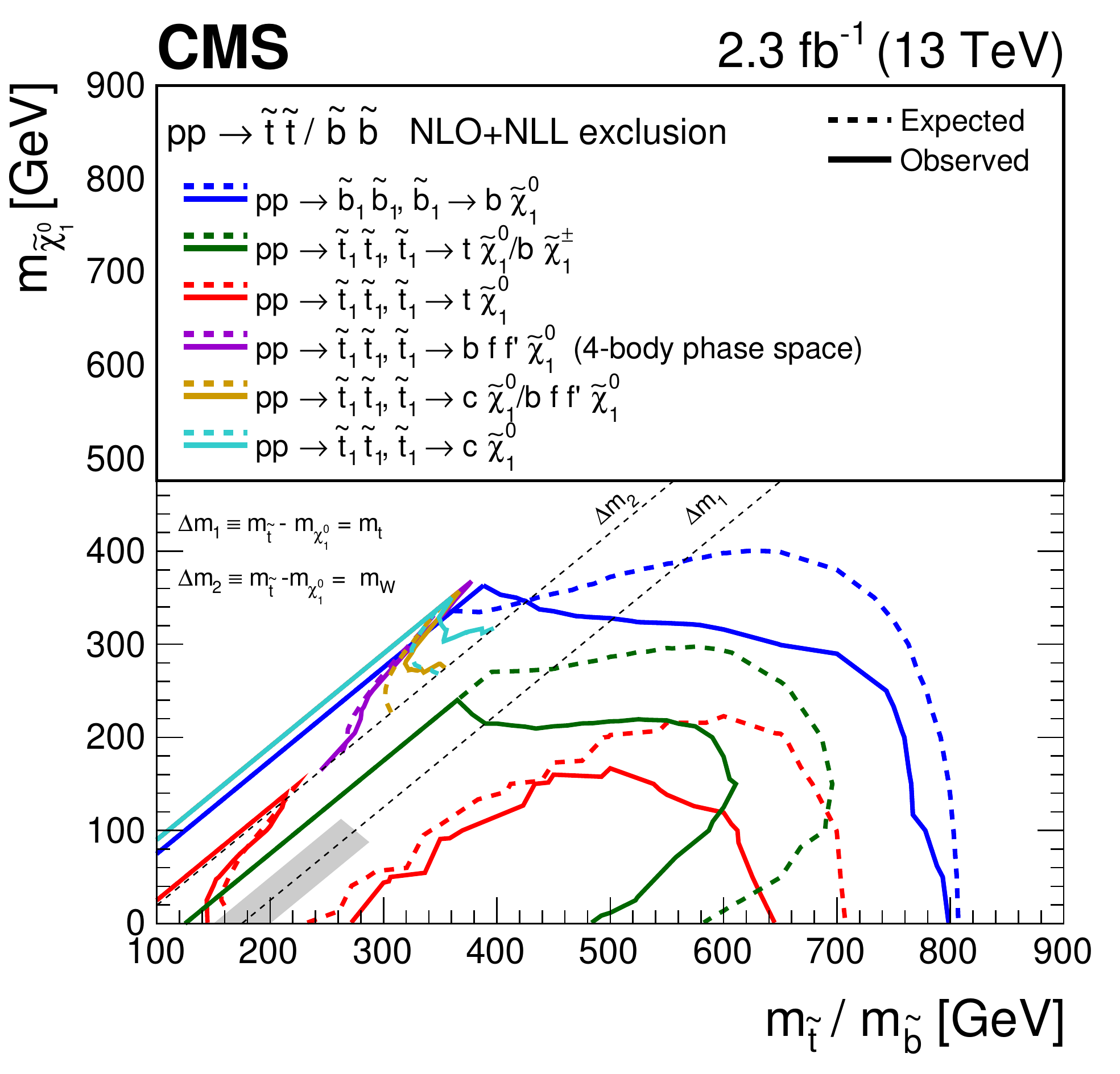}
    \caption{ Observed and expected mass exclusions at 95\% CL
      (indicated, respectively, by solid and dashed contours) for a
      number of simplified models. (\cmsLeft) Two scenarios involving the
      gluino-mediated pair production of on-shell top squarks:
      $\PSg\to \cPaqt\PSQt\to \cPaqt\cPqt\PSGczDo$ with $m_{\,\PSQt} -
      m_{\PSGczDo} = 175\GeV$ (\texttt{T5tttt\_DM175}) and
      $\PSg\to \cPaqt\PSQt\to \cPaqt\cPqc\PSGczDo$ with $m_{\,\PSQt} -
      m_{\PSGczDo} = 20\GeV$ (\texttt{T5ttcc}). Also shown, for
      comparison, is \texttt{T1tttt}. (\cmsRight) Six scenarios involving
      the direct pair production of third-generation squarks. The
      first scenario involves the pair production of bottom squarks,
      $\PSQb\to \cPqb\PSGczDo$ (\texttt{T2bb}). Two scenarios involve
      the decay of top squark pairs as follows: $\PSQt\to \cPqt\PSGczDo$
      or $\PSQt\to \cPqb\PSGcpm_1\to \cPqb\PW^*\PSGczDo$ with
      $m_{\PSGcpm_1} - m_{\PSGczDo} = 5\GeV$ and branching fractions
      $50/50\%$ (\texttt{T2tb}), or $\PSQt\to \cPqt\PSGczDo$
      (\texttt{T2tt}). The final three scenarios consider top squark
      decays under the assumption $10 < m_{\,\PSQt} - m_{\PSGczDo} <
      80\GeV$: $\PSQt\to \cPqc\PSGczDo$ (\texttt{T2cc}),
      $\PSQt\to \cPqb\PW^*\PSGczDo$ (\texttt{T2tt\_degen}), and
      $\PSQt\to \cPqc\PSGczDo$ or $\PSQt\to \cPqb\PW^*\PSGczDo$ with
      branching fractions $50/50\%$ (\texttt{T2tt\_mixed}). The grey
      shaded region denotes \texttt{T2tt} models that are not
      considered for interpretation. }
    \label{fig:limits-sms-2}
\end{figure}

Figures~\ref{fig:limits-sms-1} and~\ref{fig:limits-sms-2} summarise
the disfavoured regions of the mass parameter space for the fourteen
classes of simplified models. These regions are derived by comparing
the upper limits on the measured fiducial cross section, corrected for
the experimental $\mathcal{A}\,\varepsilon$, with the theoretical cross
sections calculated at NLO+NLL accuracy in
$\alpha_\mathrm{s}$~\cite{susynlo}. The former cross section value is
determined as a function of $m_{\PSg}$ or $m_{\PSQ}$ and
$m_{\PSGczDo}$, while the latter has a dependence solely on $m_{\PSg}$
or $m_{\PSQ}$.
The exclusion of models is evaluated using observed data counts in the
signal region (solid contours) and also expected counts based on an
Asimov data set (dashed contours).

Figure~\ref{fig:limits-sms-1} (upper) shows the excluded mass parameter
space for models that assume the gluino-mediated or direct production
of light-flavour squarks. The excluded region extends to higher masses
for the gluino-mediated production of light-flavour squarks
(\texttt{T1qqqq}), with respect to direct pair production when
assuming an eightfold degeneracy in mass (\texttt{T2qq\_8fold}), due
to a combination of a higher gluino pair production cross section and
a final state characterised by higher jet multiplicities, which are
exploited to provide better signal-to-background separation. The
excluded mass region is significantly reduced when assuming only a
single light squark (\texttt{T2qq\_1fold}), with limits weakening due
to the lower production cross section, compounded by the reduced
signal-to-background ratios achieved in the core of distributions in
the discriminating variables.

Figure~\ref{fig:limits-sms-1} (lower) shows the exclusion contours
for models that assume the gluino-mediated pair production of
off-shell third-generation squarks. For the topologies \texttt{T1tttt}
and \texttt{T1bbbb}, each gluino is assumed to undergo a three-body
decay via, respectively, an off-shell top or bottom squark to a
quark-antiquark pair of the same flavour and the $\PSGczDo$. In the
case of \texttt{T1ttbb}, each gluino is assumed to undergo a
three-body decay to an on-shell chargino, $\PSGcpm_1$, a bottom quark,
and a top antiquark. The chargino mass is defined relative to the
neutralino mass via the expression $m_{\PSGcpm_1} - m_{\PSGczDo} =
5\GeV$. The chargino decays promptly to the $\PSGczDo$ and an off-shell
W boson. The excluded mass regions differ significantly for these
topologies, primarily due to the different number of (on-shell) W
bosons in their final states, resulting in the highest $\mathcal{A}
\, \varepsilon$ for \texttt{T1bbbb} and lowest for
\texttt{T1tttt}. Further, $\mathcal{A} \, \varepsilon$ has a
strong dependence on jet multiplicity, which is highest for
\texttt{T1tttt}, due to the \bdphi variable. An additional feature for
\texttt{T1ttbb} is the weakening of the mass limit at low values of
$m_{\PSGczDo}$, when $m_{\PSGcpm_1} = m_{\PSGczDo} + 5\GeV \lesssim
m_\text{t}$. In this scenario, the $\PSGcpm_1$ (and hence $\PSGczDo$) is
not highly Lorentz boosted relative to the top quark resulting from
the three-body decay of the gluino. Hence, two $\PSGczDo$ SUSY particles do
not carry away significant \ptvecmiss, which is instead realised
through W boson decays to neutrinos and ``lost'' leptons or $\tau$
leptons that decay to neutrinos and hadrons. The observed mass limits
for these topologies are up to $\sim$2 standard deviations weaker than
the expected limits. These differences are due to upward fluctuations
in data for two contiguous bins that satisfy the requirements $\njet
\geq 5$, $\nb \geq 2$, and $\scalht > 800\GeV$. This region has the
highest sensitivity to models involving gluino production and decays
to third-generation quarks (via on- or off-shell squarks). The
observed counts are consistent with statistical fluctuations and the
events do not exhibit anomalous nonphysical behaviour. The events are
distributed in \HTmiss consistent with expectation, hence models
characterised by high values of \HTmiss, such as \texttt{T1bbbb} with
$m_{\PSg} \gg m_{\PSGczDo}$ or $m_{\PSg} \approx m_{\PSGczDo}$, are
less compatible with the data counts in this high-\njet, \nb, and
\scalht region.

Figure~\ref{fig:limits-sms-2} (upper) shows exclusion contours for
models that assume gluino pair production, with each gluino decaying
to a top quark and an on-shell top squark, the latter of which decays
to SM particles and the LSP. As discussed earlier, these
models can be considered as representations of a natural solution
to the little hierarchy problem. Two different scenarios are
considered for the decay of the top squarks. The
\texttt{T5tttt\_DM175} models assume a two-body decay to a top quark
and the $\PSGczDo$, with the top squark mass defined relative to the
$\PSGczDo$ as $m_{\PSQt} - m_{\PSGczDo} = m_\text{t}$. Models that
satisfy $m_{\PSGczDo} < 50\GeV$ are not considered here, as the
$\PSGczDo$ particles carry very little momentum.  The \texttt{T5ttcc}
models assume $m_{\PSQt} - m_{\PSGczDo} = 20\GeV$ and two-body decays
to a charm quark and the $\PSGczDo$.

Finally, Fig.~\ref{fig:limits-sms-2} (lower) shows exclusion contours
for models that assume the direct production of pairs of
third-generation squarks. For the \texttt{T2bb} models, bottom squarks
are pair produced and each decays to a bottom quark and the
$\PSGczDo$. The \texttt{T2tt} models assume top squarks are pair
produced and each is assumed to undergo a two- or three-body decay to,
respectively, a top quark and the $\PSGczDo$ when $m_{\PSQt} -
m_{\PSGczDo} > m_\text{t}$ is satisfied, or a b quark, an on-shell W
boson, and the $\PSGczDo$ for the condition $m_{\PW} < m_{\PSQt} -
m_{\PSGczDo} < m_\text{t}$. Models that satisfy $\abs{m_{\PSQt} -
  m_\text{t} - m_{\PSGczDo}} < 25\GeV$ and $m_{\PSQt} + m_{\PSGczDo} <
375\GeV$ are not considered here, as $\sigma_\text{UL}$ is a strong
function of $m_{\PSQt} - m_{\PSGczDo}$ in this low-$m_{\PSQt}$ region
due to the high levels of signal contamination found in the \mj
control region for models that resemble the \ttbar background in terms
of their topological and kinematic properties. The \texttt{T2tb}
models also assume the pair production of top squarks, with each
undergoing a two-body decay to either a top quark and the $\PSGczDo$,
or a bottom quark and the $\PSGcpm_1$, with equal branching fractions
of 50\%. As for the \texttt{T1ttbb} models, the chargino mass is
defined relative to the neutralino mass via the expression
$m_{\PSGcpm_1} - m_{\PSGczDo} = 5\GeV$, and the chargino decays promptly
to the $\PSGczDo$ and an off-shell W boson. The excluded mass regions
differ significantly for the \texttt{T2bb}, \texttt{T2tb}, and
\texttt{T2tt} topologies, in an analogous way to the \texttt{T1bbbb},
\texttt{T1ttbb}, and \texttt{T1tttt} models described above. The
difference in the mass exclusions is due primarily to the different
number of (on-shell) W bosons in the final states, which affects
$\mathcal{A} \, \varepsilon$ through the presence of leptons from
the decay of the W boson. An additional feature for \texttt{T2tb} is
the weakening of the mass limit at low values of $m_{\PSGczDo}$, when
$m_{\PSGcpm_1} = m_{\PSGczDo} + 5\GeV \lesssim m_\text{t}$. Moderately
weaker than expected mass limits are observed for all models involving
two-body decays, which is traced to mild upward fluctuations in data
for events satisfying $\njet = 2$, $\nb = 2$, and $350 < \scalht <
500\GeV$.

Figure~\ref{fig:limits-sms-2} (lower) also shows exclusion contours
for models that assume the pair production of top squarks but a
near-mass-degenerate system that satisfies $10\GeV < m_{\PSQt} -
m_{\PSGczDo} < m_\PW$. Two decays of the top squark are
considered.
The \texttt{T2cc} and \texttt{T2tt\_degen} models assume two- and
four-body decays of the top squark to, respectively, a charm quark and
the $\PSGczDo$, or to $\text{bf}\bar{\text{f}}'\PSGczDo$, where
$\text{f}$ and $\bar{\text{f}}'$ are fermions produced in the decay of
an intermediate off-shell W boson. A third class of models,
\texttt{T2tt\_mixed}, assumes both these decay modes with an equal
branching fraction of 50\%. For \texttt{T2cc}, the excluded mass
region is relatively stable as a function of the mass splitting $\dm =
m_{\PSQt} - m_{\PSGczDo}$, with $\PSQt$ masses excluded up to
400\GeV. For \texttt{T2tt\_degen}, the excluded mass region is
strongly dependent on \dm, weakening considerably for increasing
values of \dm due to the increased momentum phase space available to
leptons produced in the four-body decay. The \texttt{T2tt\_mixed}
models exhibit an intermediate behaviour. Mass limits for all three
model classes converge for the smallest mass splitting considered,
$\dm = 10\GeV$, when the SM particles from the \PSQt decay are
extremely soft and outside the experimental acceptance. An
approximately contiguous mass exclusion limit is observed across the
transition from the \texttt{T2tt\_degen} four-body to the
\texttt{T2tt} three-body decay of the $\PSQt$, as the top quark moves
on-shell. The excluded mass region weakens further as $\dm \to
m_\text{t}$.

Table~\ref{tab:simplified-models-limits} summarises the strongest
expected and observed mass limits for each class of simplified model. 

\begin{table}[!t]
  \topcaption{Summary of the mass limits obtained for the fourteen
    classes of simplified models. The limits indicate the strongest
    observed and expected (in parentheses) mass exclusions in $\PSg$,
    $\PSQ$, $\PSQb$, $\PSQt$, and $\PSGczDo$. The quoted values have
    uncertainties of $\pm$25 and $\pm$10\GeV for models involving
    the pair production of, respectively, gluinos and squarks.
  }
  \label{tab:simplified-models-limits}
  \centering
  \begin{tabular}{ lccc }
    \hline
    Model class & Parent  SUSY  & \multicolumn{2}{c}{Best mass limit [\GeVns{}]}  \\
    \cline{3-4}
                & particle & $\PSg / \PSQ / \PSQb / \PSQt$ & $\PSGczDo$\T \\ [0.5ex]
    \hline
    \texttt{T1qqqq}        & $\PSg$   & 1375 \ph(1350)                 & 875 \ph(850) \\
    \texttt{T2qq\_8fold}   & $\PSQ$    & 1150 \ph(1075)                 & 600 \ph(550) \\
    \texttt{T2qq\_1fold}   & $\PSQ$    & \ph575 \ph\ph(650)             & 275 \ph(275) \\
    \texttt{T1bbbb}        & $\PSg$   & 1575 \ph(1575)                 & 975 (1025)   \\
    \texttt{T1tttt}        & $\PSg$   & 1125 \ph(1325)                 & 475 \ph(600) \\
    \texttt{T1ttbb}        & $\PSg$   & 1375 \ph(1450)                 & 750 \ph(850) \\
    \texttt{T5tttt\_DM175}        & $\PSg$   & \ph800 \ph(1000)               & 300 \ph(450) \\
    \texttt{T5ttcc}        & $\PSg$   & 1350 \ph(1350)                 & 700 \ph(800) \\
    \texttt{T2bb}          & $\PSQb$   & \ph800 \ph\ph(800)             & 360 \ph(400) \\
    \texttt{T2tb}          & $\PSQt$   & \ph610 \ph\ph(690)             & 240 \ph(300) \\
    \texttt{T2tt} (3-body) & $\PSQt$   & \ph670 \ph\ph(720)             & 210 \ph(240) \\
    \texttt{T2tt} (2-body) & $\PSQt$   & \ph280 \ph\ph(280)             & 200 \ph(200) \\
    \texttt{T2cc}          & $\PSQt$   & \ph400 \ph\ph(350)             & 310 \ph(340) \\
    \texttt{T2tt\_degen}   & $\PSQt$   & \ph370 \ph\ph(360)             & 360 \ph(350) \\
    \texttt{T2tt\_mixed}   & $\PSQt$   & \ph360 \ph\ph(350)             & 350 \ph(340) \\ [0.5ex]
    \hline
  \end{tabular}
\end{table}

\section{Summary}
\label{sec:summary}

An inclusive search for new-physics phenomena is reported, based on
data from pp collisions at $\sqrt{s} = 13\TeV$. The data are recorded
with the CMS detector and correspond to an integrated luminosity of
$2.3 \pm 0.1 \fbinv$. The final states analysed contain one or more
jets with large transverse momenta and a significant imbalance of
transverse momentum, as expected from the production of massive
coloured SUSY particles, each decaying to SM particles and the
lightest stable, weakly-interacting, SUSY particle.

The sums of the standard model backgrounds are estimated from a
simultaneous binned likelihood fit to the observed yields for samples
of events categorised according to the number of reconstructed jets,
the number of jets identified as originating from b quarks, and the
scalar and the magnitude of the vector sums of the transverse momenta
of jets. In addition to the signal region, \mj, \mmj, and \gj control
regions are included in the likelihood fit. The observed yields are
found to be in agreement with the expected contributions from standard
model processes.  The search result is interpreted in the mass
parameter space of fourteen simplified SUSY models,
which cover scenarios that involve the gluino-mediated or direct
production of light- or heavy-flavour squarks, intermediate SUSY
particle states, as well as natural and nearly mass-degenerate
spectra.

The increase in the centre-of-mass energy of the LHC, from 8 to
13\TeV, provides a significant gain in sensitivity to heavy particle
states such as gluinos. In the case of pair-produced gluinos, each
decaying via an off-shell b squark to the b quark and the LSP, models
with masses up to $\sim$1.6 and $\sim$1.0\TeV are excluded for,
respectively, the gluino and LSP. These limits improve on those
obtained at $\sqrt{s} = 8\TeV$ by, respectively, $\sim$250 and
$\sim$300\GeV. In the case of direct pair production, models with
masses up to $\sim$800 and $\sim$350\GeV are excluded for,
respectively, the b squark and LSP. These mass limits are sensitive to
the assumptions on the squark flavour and the presence of intermediate
states such as charginos.

{\tolerance=800
Finally, a comprehensive study of nearly mass-degenerate models
involving top squark pair production is performed. The two decay modes
of the top squark are the loop-induced two-body decay to the
neutralino and one c quark, and the four-body decay to the neutralino,
one b quark, and an off-shell W boson. A third scenario is considered
in which the two modes are simultaneously open, each with a branching
fraction of 50\%. Masses of the top squark and LSP up to,
respectively, 400 and 360\GeV are excluded, depending on the decay
modes considered.
\par}

In conclusion, the analysis provides sensitivity across a large region
of the natural SUSY parameter space, as characterised by
interpretations with several simplified models. In particular, these
studies improve on existing limits for nearly mass-degenerate models
involving the production of pairs of top squarks.

\clearpage
\begin{acknowledgments}

 \hyphenation{Bundes-ministerium Forschungs-gemeinschaft Forschungs-zentren Rachada-pisek} We congratulate our colleagues in the CERN accelerator departments for the excellent performance of the LHC and thank the technical and administrative staffs at CERN and at other CMS institutes for their contributions to the success of the CMS effort. In addition, we gratefully acknowledge the computing centres and personnel of the Worldwide LHC Computing Grid for delivering so effectively the computing infrastructure essential to our analyses. Finally, we acknowledge the enduring support for the construction and operation of the LHC and the CMS detector provided by the following funding agencies: the Austrian Federal Ministry of Science, Research and Economy and the Austrian Science Fund; the Belgian Fonds de la Recherche Scientifique, and Fonds voor Wetenschappelijk Onderzoek; the Brazilian Funding Agencies (CNPq, CAPES, FAPERJ, and FAPESP); the Bulgarian Ministry of Education and Science; CERN; the Chinese Academy of Sciences, Ministry of Science and Technology, and National Natural Science Foundation of China; the Colombian Funding Agency (COLCIENCIAS); the Croatian Ministry of Science, Education and Sport, and the Croatian Science Foundation; the Research Promotion Foundation, Cyprus; the Secretariat for Higher Education, Science, Technology and Innovation, Ecuador; the Ministry of Education and Research, Estonian Research Council via IUT23-4 and IUT23-6 and European Regional Development Fund, Estonia; the Academy of Finland, Finnish Ministry of Education and Culture, and Helsinki Institute of Physics; the Institut National de Physique Nucl\'eaire et de Physique des Particules~/~CNRS, and Commissariat \`a l'\'Energie Atomique et aux \'Energies Alternatives~/~CEA, France; the Bundesministerium f\"ur Bildung und Forschung, Deutsche Forschungsgemeinschaft, and Helmholtz-Gemeinschaft Deutscher Forschungszentren, Germany; the General Secretariat for Research and Technology, Greece; the National Scientific Research Foundation, and National Innovation Office, Hungary; the Department of Atomic Energy and the Department of Science and Technology, India; the Institute for Studies in Theoretical Physics and Mathematics, Iran; the Science Foundation, Ireland; the Istituto Nazionale di Fisica Nucleare, Italy; the Ministry of Science, ICT and Future Planning, and National Research Foundation (NRF), Republic of Korea; the Lithuanian Academy of Sciences; the Ministry of Education, and University of Malaya (Malaysia); the Mexican Funding Agencies (BUAP, CINVESTAV, CONACYT, LNS, SEP, and UASLP-FAI); the Ministry of Business, Innovation and Employment, New Zealand; the Pakistan Atomic Energy Commission; the Ministry of Science and Higher Education and the National Science Centre, Poland; the Funda\c{c}\~ao para a Ci\^encia e a Tecnologia, Portugal; JINR, Dubna; the Ministry of Education and Science of the Russian Federation, the Federal Agency of Atomic Energy of the Russian Federation, Russian Academy of Sciences, and the Russian Foundation for Basic Research; the Ministry of Education, Science and Technological Development of Serbia; the Secretar\'{\i}a de Estado de Investigaci\'on, Desarrollo e Innovaci\'on and Programa Consolider-Ingenio 2010, Spain; the Swiss Funding Agencies (ETH Board, ETH Zurich, PSI, SNF, UniZH, Canton Zurich, and SER); the Ministry of Science and Technology, Taipei; the Thailand Center of Excellence in Physics, the Institute for the Promotion of Teaching Science and Technology of Thailand, Special Task Force for Activating Research and the National Science and Technology Development Agency of Thailand; the Scientific and Technical Research Council of Turkey, and Turkish Atomic Energy Authority; the National Academy of Sciences of Ukraine, and State Fund for Fundamental Researches, Ukraine; the Science and Technology Facilities Council, UK; the US Department of Energy, and the US National Science Foundation.

Individuals have received support from the Marie-Curie programme and the European Research Council and EPLANET (European Union); the Leventis Foundation; the A. P. Sloan Foundation; the Alexander von Humboldt Foundation; the Belgian Federal Science Policy Office; the Fonds pour la Formation \`a la Recherche dans l'Industrie et dans l'Agriculture (FRIA-Belgium); the Agentschap voor Innovatie door Wetenschap en Technologie (IWT-Belgium); the Ministry of Education, Youth and Sports (MEYS) of the Czech Republic; the Council of Science and Industrial Research, India; the HOMING PLUS programme of the Foundation for Polish Science, cofinanced from European Union, Regional Development Fund, the Mobility Plus programme of the Ministry of Science and Higher Education, the National Science Center (Poland), contracts Harmonia 2014/14/M/ST2/00428, Opus 2013/11/B/ST2/04202, 2014/13/B/ST2/02543 and 2014/15/B/ST2/03998, Sonata-bis 2012/07/E/ST2/01406; the Thalis and Aristeia programmes cofinanced by EU-ESF and the Greek NSRF; the National Priorities Research Program by Qatar National Research Fund; the Programa Clar\'in-COFUND del Principado de Asturias; the Rachadapisek Sompot Fund for Postdoctoral Fellowship, Chulalongkorn University and the Chulalongkorn Academic into Its 2nd Century Project Advancement Project (Thailand); and the Welch Foundation, contract C-1845.
\end{acknowledgments}

\bibliography{auto_generated}
\cleardoublepage \appendix\section{The CMS Collaboration \label{app:collab}}\begin{sloppypar}\hyphenpenalty=5000\widowpenalty=500\clubpenalty=5000\textbf{Yerevan Physics Institute,  Yerevan,  Armenia}\\*[0pt]
V.~Khachatryan, A.M.~Sirunyan, A.~Tumasyan
\vskip\cmsinstskip
\textbf{Institut f\"{u}r Hochenergiephysik,  Wien,  Austria}\\*[0pt]
W.~Adam, E.~Asilar, T.~Bergauer, J.~Brandstetter, E.~Brondolin, M.~Dragicevic, J.~Er\"{o}, M.~Flechl, M.~Friedl, R.~Fr\"{u}hwirth\cmsAuthorMark{1}, V.M.~Ghete, C.~Hartl, N.~H\"{o}rmann, J.~Hrubec, M.~Jeitler\cmsAuthorMark{1}, A.~K\"{o}nig, I.~Kr\"{a}tschmer, D.~Liko, T.~Matsushita, I.~Mikulec, D.~Rabady, N.~Rad, B.~Rahbaran, H.~Rohringer, J.~Schieck\cmsAuthorMark{1}, J.~Strauss, W.~Waltenberger, C.-E.~Wulz\cmsAuthorMark{1}
% ...
\vskip\cmsinstskip
% ...
1:~~Also at Vienna University of Technology, Vienna, Austria\\
% ...

\end{sloppypar}
\end{document}